%% file: bbkk14.tex
\documentclass{hkpaper}
\pdfoutput=1
\usepackage{hyperref}
\usepackage{hkthm}
\input{macros}

\usepackage{savethm} 
\settheoremname{lemma}{Lemma}
\settheoremname{proposition}{Proposition}
\settheoremname{definition}{Definition}
\settheoremname{theorem}{Theorem}
\settheoremname{example}{Example}

\usepackage[utf8]{inputenc}
\usepackage[USenglish]{babel}

\usepackage{amsmath}

\theoremstyle{hkplain}
\newtheorem{theorem_apx}{Theorem}[subsection]
\newtheorem{proposition_apx}[theorem_apx]{Proposition}
\newtheorem{lemma_apx}[theorem_apx]{Lemma}

\theoremstyle{hkdefinition}
\newtheorem{definition_apx}[theorem_apx]{Definition}

\title{Behavioral Metrics via Functor Lifting\thanks{This is an extended version of \cite{BBKK14}.}}
\author[1]{Paolo Baldan}
\author[2]{Filippo Bonchi}
\author[3]{Henning Kerstan}
\author[3]{Barbara~König}
\affil[1]{Dipartimento di Matematica, Università di Padova, Italy\\\href{mailto:baldan@math.unipd.it}{\texttt{baldan@math.unipd.it}}}%
\affil[2]{CNRS, ENS Lyon, Université de Lyon, France\\\href{mailto: filippo.bonchi@ens-lyon.fr}{\texttt{filippo.bonchi@ens-lyon.fr}}}%
\affil[3]{Universität Duisburg-Essen, Germany\\\href{mailto:henning.kerstan@uni-due.de}{\texttt{henning.kerstan@uni-due.de}}, \href{mailto:barbara_koenig@uni-due.de}{\texttt{barbara\_koenig@uni-due.de}}}

\keywords{behavioral metric, functor lifting, pseudometric, coalgebra}% mandatory: Please provide 1-5 keywords
\begin{document}
\maketitle 
\input{main}
\appendix
\include{apx-fibrations}

\include{apx-proofs}
\end{document}

%% file: macros.tex
\renewcommand{\phi}{\varphi}
\renewcommand{\epsilon}{\varepsilon}

\newcommand{\PMet}{\mathbf{PMet}}
\newcommand{\Set}{\mathbf{Set}}
\newcommand{\Coalg}[1]{\mathbf{Coalg}(#1)}
\newcommand{\one}{\mathbf{1}}
\newcommand{\Id}{\mathrm{Id}}
\newcommand{\id}{\mathrm{id}}
\newcommand{\nonexpansiveTo}{\raisebox{-1.5pt}{\ensuremath{\,\mathop{\overset{1}{\to}}\,}}}

\newcommand{\preal}{\mathbb{R}_0^+}
\newcommand{\prealinf} {\ensuremath{[0,\infty]}}
\newcommand{\reals}{{[0,\top]}}
\newcommand{\N}{\mathbb{N}}
\newcommand{\set}[1]{\left\{#1\right\}}

\newcommand{\ExpectedValue}[2]{\ensuremath{\mathbb{E}_{#1}[{#2}]}} %expected value
\newcommand{\ev}{\ensuremath{ev}} % evaluation function
\newcommand{\Kantorovich}[2]{\ensuremath{#2^{\,\uparrow {#1}}}} % \Kantorovich{Functor}{Metric}
\newcommand{\KantorovichMulti}[3]{\ensuremath{{#2}^{\,\uparrow {#1}}_{#3}}}
\newcommand{\LiftedFunctor}[1]{\ensuremath{\overline{#1}}}
\newcommand{\EvaluationFunctor}[1]{\ensuremath{\widetilde{#1}}}
\newcommand{\Wasserstein}[2]{\ensuremath{#2^{\,\downarrow {#1}}}} % \Wasserstein{Functor}{Metric}
\newcommand{\WassersteinMulti}[3]{\ensuremath{{#2}^{\,\downarrow {#1}}_{#3}}}
\newcommand{\LiftedMetric}[2]{\ensuremath{#2^{#1}}}
\newcommand{\Couplings}[1]{\Gamma_{#1}}
\newcommand{\PowersetFinite}{\ensuremath{\mathcal{P}_{\!f\!i\!n}}}
\newcommand{\Distributions}{\ensuremath{\mathcal{D}}}

\newcommand{\refbyname}[2]{\hyperref[#1]{#2~\ref*{#1}}}
\newcommand{\definitionref}[1]{\refbyname{#1}{Definition}}
\newcommand{\lemmaref}[1]{\refbyname{#1}{Lemma}}
\newcommand{\exampleref}[1]{\refbyname{#1}{Example}}
\newcommand{\propositionref}[1]{\refbyname{#1}{Proposition}}

\newcommand{\figureref}[1]{\refbyname{#1}{Figure}}
\newcommand{\apxpropositionref}[1]{\refbyname{#1}{Proposition}}
\newcommand{\theoremref}[1]{\refbyname{#1}{Theorem}}
\newcommand{\appendixref}[1]{\refbyname{#1}{Appendix}}

\newcommand{\lbbd}{\mathopen{[\![}}
\newcommand{\rbbd}{\mathclose{]\!]}}
\newcommand{\final}[1]{\lbbd #1 \rbbd}
\newcommand{\finalO}{\final{\_}}

\usepackage[colorinlistoftodos, textsize=footnotesize,color=orange!50]{todonotes}

\usepackage{tikz}
\usetikzlibrary{matrix}
\usetikzlibrary{arrows}
\usetikzlibrary{automata}

%% file: main.tex
%!TEX root = paper-129.tex

\begin{abstract}
\noindent We study behavioral metrics in an abstract coalgebraic setting. Given a coalgebra $\alpha\colon X \to FX$ in $\Set$, where the functor $F$ specifies the branching type, we define a framework for deriving pseudometrics on $X$ which measure the behavioral distance of states.

A first crucial step is the lifting of the functor $F$ on $\Set$ to a functor $\overline{F}$ in the category $\PMet$ of pseudometric spaces. We present two different approaches which can be viewed as generalizations of the Kantorovich and Wasserstein pseudometrics for probability measures. We show that the pseudometrics provided by the two approaches coincide on several natural examples, but in general they differ. 

Then a final coalgebra for $F$ in $\Set$ can be endowed with a behavioral distance resulting as the smallest solution of a fixed-point equation, yielding the final $\overline{F}$-coalgebra in $\PMet$. The same technique, applied to an arbitrary coalgebra $\alpha\colon X \to FX$ in $\Set$,  provides the behavioral distance on $X$. Under some constraints we can prove that two states are at distance $0$ if and only if they are behaviorally equivalent.
\end{abstract}

\section{Introduction}
Increasingly, modelling formalisms are equipped with quantitative
information, such as probability, time or weight. Such quantitative
information should be taken into account when reasoning about
behavioral equivalence of system states, such as bisimilarity.  In
this setting the appropriate notion is not necessarily equivalence,
but a behavioral metric that measures the distance of the behavior of
two states. In a quantitative setting, it is often unreasonable to
assume that two states have exactly the same behavior, but it makes
sense to express that their behavior differs by some (small) value
$\varepsilon$. 

The above considerations led to the study of behavioral metrics which
aims at quantifying the the distance between the behavior of
states. Since different states can have exactly the same behavior it
is quite natural to consider \emph{pseudometrics}, which allow
different elements to be at zero distance.

Earlier contributions defined behavioral metrics in the setting of probabilistic systems \cite{DGJP04,vBW06} and of metric transition systems
\cite{afs:linear-branching-metrics}.
Our aim is to generalize these ideas and to study behavioral metrics
in a general coalgebraic setting. The theory of coalgebra \cite{Rut00} is nowadays a well-established tool for defining and
reasoning about various state based transition systems such as
deterministic, nondeterministic, weighted or probabilistic automata.
Hence, it is the appropriate setting to ask and answer general
questions about behavioral metrics.

\begin{itemize}
\item \emph{How can we define behavioral metrics for transition
    systems with different branching types?} We provide a coalgebraic
  framework in the category of pseudometric spaces $\PMet$ that allows
  to define and reason about such metrics.
\item \emph{Are the behavioral metrics canonical in some way?} We
  provide a natural way to define metrics by lifting functors from
  $\Set$ to the category of pseudometric spaces.  In fact, we study
  two liftings: the Kantorovich and the Wasserstein lifting and
  observe that they coincide in many cases. This provides us with a
  notion of canonicity and justification for the choice of metrics.
\item \emph{Does the measurement of distances affect behavioral
    equivalence?} If we start by considering coalgebras in $\PMet$
  (as, e.g., in \cite{vBW06}), it is not entirely clear
  a priori whether the richer categorical structure influences the
  notion of behavioral equivalence. In our setting we start with
  coalgebras in $\Set$ and put distance measurements ``on top'',
  showing that, under some mild constraints, the original notion of
  behavioral equivalence is not compromised, in the sense that two
  states are behaviorally equivalent iff their distance is $0$.
\item \emph{Are there generic algorithms to compute metrics?}
  Coalgebra is a valuable tool to define generic methods that can be
  instantiated to concrete cases in order to obtain prototype
  algorithms. In our case we give a (high-level) procedure for
  computing behavioral distances on a given coalgebra, based
  on determining the smallest solution of a fixed-point equation.
\end{itemize}

\noindent A central contribution of this paper is the lifting of a
functor $F$ from $\Set$ to $\PMet$. Given a pseudometric space
$(X,d)$, the goal is to define a suitable pseudometric on $FX$.  Such
liftings of metrics have been extensively studied in transportation
theory \cite{v:optimal-transport}, e.g.\ for the case of the
(discrete) probability distribution functor, which comes with a nice
analogy: assume several cities (with fixed distances between them) and
two probability distributions $s,t$ on cities, representing supply and
demand (in units of mass).  The distance between $s,t$ can be measured
in two ways: the first is to set up an optimal transportation plan
with minimal costs (in the following also called coupling) to
transport goods from cities with excess supply to cities with excess
demand. The cost of transport is determined by the product of mass and
distance. In this way we obtain the Wasserstein distance.  A different
view is to imagine a logistics firm that is commissioned to handle the
transport.  It sets prices for each city and buys and sells for this
price at every location.  However, it has to ensure that the price
function is nonexpansive, i.e., the difference of prices between two
cities is smaller than the distance of the cities, otherwise it will
not be worthwhile to outsource this task. This firm will attempt to
maximize its profit, which can be considered as the Kantorovich
distance of $s,t$. The Kantorovich-Rubinstein duality informs us that
these two views lead to the exactly same result, a very good argument
for the canonicity of this notion of distance.

It is our observation that these two notions of distance lifting can
analogously be defined for arbitrary functors $F$, leading to a rich
general theory.  The lifting has an evaluation function as
parameter. As concrete examples, besides the probability distribution
functor, we study the (finite) powerset functor (resulting in the
Hausdorff metric) and the coproduct and product bifunctors. In the
case of the product bifunctor we consider different evaluation
functions, each leading to a well-known product metric. The
Kantorovich-Rubinstein duality holds for these functors, but it does
not hold in general (we provide a counterexample).

After discussing functor liftings, we define coalgebraic behavioral
pseudometrics and answer the questions above. Specifically we
show how to compute distances on the final coalgebra as well as on
arbitrary coalgebras via fixed-point iteration and we prove that the
pseudometric obtained on the final coalgebra is indeed a metric. 
In \appendixref{app:fibrations} we discuss a fibrational perspective on our work and we compare with~\cite{HasuoCKJ13}.  All proofs for our results are in \appendixref{sec:proofs}.

\section{Preliminaries, Notation \& Evaluation Functions}
\label{sec:preliminaries}

We assume that the reader is familiar with the basic notions of category theory, especially with the definitions of functor, product, coproduct and weak pullbacks.

For a function $f\colon X \to Y$ and sets $A \subseteq X$, $B \subseteq Y$ we write $f[A] := \set{f(a) \mid a \in A}$ for the \emph{image} of $A$ and $f^{-1}[B]= \set{a \in A \mid f(x) \in B}$ for the \emph{preimage} of $B$. If $Y \subseteq [0,\infty]$ and $f,g\colon X \to Y$ are functions we write $f \leq g$ when $\forall x \in X:f(x)\leq g(x)$.

Given a natural number $n \in \N$ and a family $(X_i)_{i = 1}^n$ of sets $X_i$ we denote the projections of the (cartesian) product of the $X_i$ by $\pi_i^n\colon \prod_{i=1}^n X_i \to X_i,$ or just by $\pi_i$ if $n$ is clear from the context. For a source $(f_i\colon X \to X_i)_{i = 1}^n$ we denote the unique mediating arrow to the product by $\langle f_1,\dots,f_n\rangle \colon X \to \prod_{i=1}^{n}X_i$. Similarly, given a family of arrows $(f_i\colon X_i \to Y_i)_{i = 1}^n$, we write $f_1 \times \dots \times f_n = \langle f_1 \circ \pi_1,\dots,f_n \circ \pi_n \rangle\colon \prod_{i=1}^n X_i \to \prod_{i=1}^n Y_i$.

We quickly recap the basic ideas of coalgebras. Let $F$ be an endofunctor on the category $\Set$ of sets and functions. An $F$-coalgebra is just a function $\alpha \colon X \to FX$. Given another $F$-coalgebra $\beta\colon Y \to FY$ a coalgebra
homomorphism from $\alpha$ to $\beta$ is a function $f \colon A \to B$
such that $\beta \circ f = Ff \circ \alpha$. We call an $F$-coalgebra
$\kappa\colon \Omega \to F\Omega$ \emph{final} if for any other
coalgebra $\alpha \colon X \to FX$ there is a unique coalgebra
homomorphism $\finalO \colon X \to \Omega$. The final coalgebra need
not exist but if it does it is unique up to isomorphism. It can be
considered as the universe of all possible behaviors. If we have an
endofunctor $F$ such that a final coalgebra $\kappa\colon \Omega \to
F\Omega$ exists then for any coalgebra $\alpha\colon X \to FX$ two
states $x_1, x_2 \in X$ are said to be \emph{behaviorally equivalent}
if and only if $\final{x_1} = \final{x_2}$.

We now introduce some preliminaries about (pseudo)metric spaces. Our
(pseudo)metrics assume values in a closed interval $[0,\top]$, where
$\top \in (0,\infty]$ is a fixed maximal element (for our examples we
will use $\top = 1$ or $\top=\infty$). In this way the set of
(pseudo)metrics over a fixed set with pointwise order is a complete lattice (since $\reals$ is) and
the resulting category of pseudometric spaces is complete and
cocomplete.

\begin{definition}[Pseudometric, Pseudometric Space]
\label{def:pseudometric}
Given a set $X$, a \emph{pseudometric on $X$} is a function $d\colon X\times X\to \reals$ such that for all $x,y,z\in X$, the following axioms hold: $d(x,x) = 0$ (\emph{reflexivity}), $d(x,y) = d(y,x)$ (\emph{symmetry}), $d(x,z) \le d(x,y)+d(y,z)$ (\emph{triangle inequality}). If additionally $d(x,y)=0$ implies $x=y$, $d$ is called a \emph{metric}. A \emph{(pseudo)metric space} is a pair $(X,d)$ where $X$ is a set and $d$ is a (pseudo)metric on $X$.
\end{definition}

\noindent By $d_e\colon \reals^2\to \reals$ we denote the ordinary Euclidean distance on $\reals$, i.e., $d_e(x,y) = |x-y|$ for $x,y\in\reals\setminus\{\infty\}$, and -- where appropriate -- $d_e(x,\infty) = \infty$ if $x\neq \infty$ and $d_e(\infty,\infty) = 0$. Addition is defined in the usual way, in particular $x + \infty = \infty$ for $x\in\prealinf$. 

Hereafter, we only consider those functions between two pseudometric spaces that do not increase the distances.

\begin{definition}[Nonexpansive Function, Isometry]
\label{def:nonexpansive-fct}
Let $(X,d_X)$, $(Y,d_Y)$ be pseudometric spaces. A function $f\colon X\to Y$ is called \emph{nonexpansive} if $d_Y \circ (f \times f) \le d_X$. In this case we write\footnote{The nonexpansive functions correspond exactly to the Lipschitz-continuous functions with Lipschitz constant $L \leq 1$, which is the reason for writing $f \colon (X, d_X) \nonexpansiveTo (Y, d_Y)$.} $f \colon (X, d_X) \nonexpansiveTo (Y, d_Y)$. If equality holds, $f$ is called an \emph{isometry}. 
\end{definition}

\noindent For our purposes it will turn out to be useful to consider the following alternative characterization of the triangle inequality using the concept of nonexpansive functions.

\begin{savetheorem}{lemma}{lem:alt-char-triangle}
A symmetric function $d\colon X^2\to \reals$ with $d(x,x) = 0$ for all $x \in X$ satisfies the triangle inequality iff for all $x \in X$ the function $d(x,\_)\colon X \to \reals$ is nonexpansive.
\end{savetheorem}

\noindent As stated before, our definition of a pseudometric gives rise to a suitably rich category.

\begin{definition}[Category of Pseudometric Spaces]
\label{def:category-metric}
For a fixed $\top \in (0,\infty]$ we denote by $\PMet$ the category of all pseudometric spaces and nonexpansive functions. 
\end{definition}

\noindent This category is complete and cocomplete (see \apxpropositionref{prop:PMetcomplete}) and, in particular, it has products and coproducts as we will see in Examples~\ref{exa:product-bifunctor} and~\ref{exa:coproduct-bifunctor}. We now introduce two motivating examples borrowed from~\cite{vBW06} and~\cite{afs:linear-branching-metrics}.

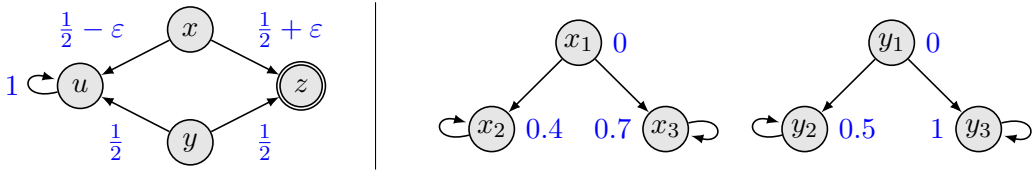
\begin{figure}
\begin{minipage}{5cm}
\begin{tikzpicture}[->,auto,node distance=1cm,semithick,>=latex]
\tikzstyle{every state}=[fill=black!10,text=black,inner sep=0.5pt, minimum size=17pt]
\tikzstyle{every node}=[text=blue]
% nodes
\node[state] (x) {$x$};
\node (center) [below=0.3cm of x] {};
\node[state] (u) [left=1cm of center]{$u$};
\node[state,accepting] (z) [right=1cm of center] {$z$};
\node[state] (y) [below=0.3 cm of center] {$y$};
% paths
\path (x) edge node[above left] {$\frac{1}{2}-\epsilon$} (u);
\path (x) edge node[above right] {$\frac{1}{2}+\epsilon$} (z);
\path (y) edge node[below left] {$\frac{1}{2}$} (u);
\path (y) edge node[below right] {$\frac{1}{2}$} (z);
\path (u) edge[loop left] node[left] {$1$} (u);
\end{tikzpicture}
\end{minipage}\vrule\hspace{0.6cm}
\begin{minipage}{8cm}
\begin{tikzpicture}[->,auto,node distance=1cm,semithick,>=latex]
\tikzstyle{every state}=[fill=black!10,text=black,inner sep=0.5pt, minimum size=17pt]
\tikzstyle{every node}=[text=blue]
% nodes system 1
\node[state] (x1) {$x_1$};
\node (p1) [right=0mm of x1] {$0$};
\node[state] (x2) [below left=of x1] {$x_2$};
\node (p2) [right=0mm of x2] {$0.4$};
\node[state] (x3) [below right=of x1] {$x_3$};
\node (p3) [left=0mm of x3] {$0.7$};
% nodes system 2
\node[state] (y1) [right=3.5cm of x1] {$y_1$};
\node (q1) [right=0mm of y1] {$0$};
\node[state] (y2) [below left=of y1] {$y_2$};
\node (q2) [right=0mm of y2] {$0.5$};
\node[state] (y3) [below right=of y1] {$y_3$};
\node (q3) [left=0mm of y3] {$1$};
% paths system 1
\path (x1) edge (x2);
\path (x1) edge (x3);
\path (x2) edge[loop left] (x2);
\path (x3) edge[loop right] (x3);
% paths system 2
\path (y1) edge (y2);
\path (y1) edge (y3);
\path (y2) edge[loop left] (y2);
\path (y3) edge[loop right] (y3);
\end{tikzpicture}
\end{minipage}
\caption{A probabilistic transition system (left) and a metric transition system (right).}
\label{fig:example-ts}
\end{figure}

\begin{example}[Probabilistic Transition Systems and Behavioral Distance]
\label{ex:probabilistic-1}
We regard probabilistic transition systems as coalgebras of the form $\alpha\colon X\to \Distributions (X+\mathbf{1})$, where $\Distributions $ is the probability distribution functor (with finite support) which maps a set $X$ to the set $\Distributions X = \{P\colon X\to [0,1]\mid \sum_{x\in X} P(x) = 1, P \text{ has finite support}\}$ and a function $f \colon X \to Y$ to the function $\Distributions f \colon \Distributions X \to \Distributions Y, P \mapsto \lambda y.\sum_{x \in   f^{-1}[\set{y}]}P(x)$. Here $\alpha(x)(y)$, for $x,y\in X$, denotes the probability of a transition from a state $x$ to $y$ and $\alpha(x)(\checkmark)$ stands for the probability of terminating from $x$ (we use $\checkmark$ for the single element of the set $\mathbf{1}$).

In \cite{vBW06} a metric for the continuous version of these systems is introduced, by considering a discount factor $c \in (0,1)$. In the discrete case we obtain the behavioral distance $d\colon X^2\to [0,1]$, defined as the least solution of the equation $d(x,y) = \overline{d}(\alpha(x),\alpha(y))$, where $x,y\in X$ and $\overline{d}\colon (\Distributions (X+\mathbf{1}))^2\to [0,1]$ is defined in two steps: First, $\hat{d}\colon (X+\mathbf{1})^2\to [0,1]$ is defined as $\hat{d}(x,y) = c\cdot d(x,y)$ if $x,y\in X$, $\hat{d}(\checkmark,\checkmark) = 0$ and $1$ otherwise. Then, for all $P_1,P_2\in \Distributions (X+\mathbf{1})$, $\overline{d}(P_1,P_2)$ is defined as the supremum of all values $\sum_{x \in X+\mathbf{1}} f(x)\cdot \big(P_1(x) - P_2(x)\big)$, with $f \colon (X+\mathbf{1},\hat{d}) \nonexpansiveTo ([0,1],d_e)$ being an arbitrary nonexpansive function.  As we will further discuss in \exampleref{exa:probability-distribution-functor}, $\overline{d}$ is the Kantorovich pseudometric given by the space $(X+\mathbf{1},\hat{d})$. 

We consider a concrete example from \cite{vBW06}, illustrated on the left of \figureref{fig:example-ts}. The behavioral distance of $u$ and $z$ is $d(u,z) = 1$ and hence $d(x,y) = c\cdot \epsilon$.
\end{example}

\begin{example}[Metric Transition Systems and Propositional Distances]
  \label{ex:metric-ts-1}
  We give another example based on the notions of
  \cite{afs:linear-branching-metrics}. A finite set $\Sigma =
  \{r_1,\dots,r_n\}$ of propositions is given and each proposition
  $r\in\Sigma$ is associated with a pseudometric space $(M_r,d_r)$. A
  valuation $u$ is a function with domain $\Sigma$ that assigns to
  each $r\in\Sigma$ an element of $M_r$. We denote the set of all
  valuations by $\mathcal{U}[\Sigma]$. A metric transition system is a
  tuple $M = (S,\tau,\Sigma,[\cdot])$ with a set $S$ of states, a
  transition relation $\tau\subseteq S\times S$, a finite set $\Sigma$
  of propositions and a valuation $[s]$ for each state $s\in S$. We
  write $\tau(s)$ for $\{s'\in S\mid (s,s')\in\tau\}$ and require that
  $\tau(s)$ is finite.

  In \cite{afs:linear-branching-metrics} the propositional distance
  between two valuations is given by $\LiftedFunctor{\mathit{pd}}(u,v)
  = \linebreak \max_{r\in\Sigma} d_r(u(r),v(r))$ for
  $u,v\in\mathcal{U}[\Sigma]$.  The (undirected) branching distance
  $d\colon S\times S\to\preal$ is defined as the smallest fixed-point
  of the following equation, where $s,t\in S$:
  \begin{equation}\label{eq:Hausd}
  d(s,t) = \max\{\LiftedFunctor{\mathit{pd}}([s],[t]),
  \max_{s'\in\tau(s)}\min_{t'\in\tau(t)}
  d(s',t'),\max_{t'\in\tau(t)}\min_{s'\in\tau(s)} d(s',t') \} 
  \end{equation} 
  Note that, apart from the first argument, this coincides with the Hausdorff distance.

  We consider an example which appears similarly
  in~\cite{afs:linear-branching-metrics} (see
  Figure~\ref{fig:example-ts}, right) with a single proposition $r\in
  \Sigma$, where $M_r = [0,1]$ is equipped with the Euclidean distance
  $d_e$. According to~\eqref{eq:Hausd}, $d(x_1,y_1)$ equals the
  Hausdorff distance of the reals associated with the sets of
  successors, which is~$0.3$ (since this is the maximal distance of
  any successor to the closest successor in the other set of
  successors, here: the distance from~$y_3$ to~$x_3$).

  In order to model such transition systems as coalgebras we define
  the following $n$-ary auxiliary functor: $G(X_1,\dots,X_n) =
  \{u\colon \Sigma\to X_1+\dots+X_n\mid u(r_i) \in X_i \}$.  Then
  coalgebras are of the form $\alpha\colon S\to
  G(M_{r_1},\dots,M_{r_n})\times \PowersetFinite(S)$, where
  $\PowersetFinite$ is the finite powerset functor and
  $\alpha(s) = ([s],\tau(s))$. As we will see later in
  \exampleref{ex:metric-ts-2}, the right-hand side of
  \eqref{eq:Hausd} can be seen as lifting a metric $d$ on $X$ to a
  metric on $G(M_{r_1},\dots,M_{r_n})\times
  \PowersetFinite(X)$.
\end{example}

\noindent Generalizing from the examples, we now establish a general framework for deriving such behavioral distances. In both cases, the crucial step is to find, for a functor $F$, a way to lift a pseudometric on $X$ to a pseudometric on $FX$. Based on this, one can set up a fixed-point equation and define behavioral distance as its smallest solution. Hence, in the next sections we describe how to lift an endofunctor $F$ on $\Set$ to an endofunctor on $\PMet$.

\begin{definition}[Lifting]
\label{def:lifting}
Let $U\colon \PMet \to \Set$ be the forgetful functor which maps every pseudometric space to its underlying set. A functor $\LiftedFunctor{F}\colon\PMet\to\PMet$ is called a \emph{lifting} of a functor $F\colon \Set \to \Set$ if it satisfies $U\circ \LiftedFunctor{F} = F\circ U$.
\end{definition}

\noindent It is not difficult to prove that such a lifting is always
monotone on pseudometrics over a common set, i.e. for any two
pseudometrics $d_1 \leq d_2$ on the same set $X$, we also have
$\LiftedMetric{F}{d_1} \leq \LiftedMetric{F}{d_2}$ where
$\LiftedMetric{F}{d_i}$ are the pseudometrics on $FX$ obtained by
applying $\LiftedFunctor{F}$ to $(X, d_i)$ (see \apxpropositionref{prop:monotone}). 
\noindent Similarly to predicate lifting of coalgebraic modal logic \cite{DBLP:journals/tcs/Schroder08}, liftings on $\PMet$ can be conveniently defined via an evaluation function.

\begin{definition}[Evaluation Function \& Evaluation Functor]
\label{def:evfct} 
Let $F$ be an endofunctor on $\Set$. An \emph{evaluation function} for $F$ is a function $\ev_F\colon F\reals \to \reals$. Given such a function, we define the \emph{evaluation functor} to be the endofunctor $\EvaluationFunctor{F}$ on $\Set/\reals$, the slice category\footnote{The slice category $\Set/\reals$ has as objects all functions $g\colon X\to\reals$ where $X$ is an arbitrary set. Given $g$ as before and $h\colon Y \to \reals$, an arrow from $g$ to $h$ is a function $f\colon X \to Y$ satisfying $h \circ f = g$.} over $\reals$, via $\EvaluationFunctor{F}(g) = \ev_F\circ Fg$ for all $g \in \Set/\reals$. On arrows $\EvaluationFunctor{F}$ is defined as $F$. 
\end{definition}

\section{Lifting Functors to Pseudometric Spaces à la Kantorovich}
\label{sec:lifting}
Let us now consider an endofunctor $F$ on $\Set$ with an evaluation function $\ev_F$. Given a pseudometric space $(X,d)$, our first approach will be to take the smallest possible pseudometric $d^F$ on $FX$ such that, for all nonexpansive functions $f\colon (X,d) \nonexpansiveTo (\reals,d_e)$, also $\EvaluationFunctor{F}f\colon (FX,d^F)\nonexpansiveTo (\reals,d_e)$ is nonexpansive again, i.e. we want to ensure that for all $t_1,t_2 \in FX$ we have $d_e(\EvaluationFunctor{F}f(t_1), \EvaluationFunctor{F}f(t_2))\leq d^F(t_1,t_2)$. This idea immediately leads us to the next definition.

\begin{savetheorem}[Kantorovich Pseudometric \& Kantorovich Lifting]{definition}{def:kantorovich}
Let $F\colon \Set \to \Set$ be a functor with an evaluation function $\ev_F$. For every pseudometric space $(X,d)$ the \emph{Kantorovich pseudometric} on $FX$ is the function $\Kantorovich{F}{d}\colon FX\times FX\to \reals$, where for all $t_1,t_2 \in FX$:
\begin{align*}
	\Kantorovich{F}{d}(t_1,t_2) := \sup \{d_e(\EvaluationFunctor{F}f(t_1),\EvaluationFunctor{F}f(t_2)) \mid f\colon (X,d) \nonexpansiveTo (\reals,d_e)\}\,.
\end{align*}
The \emph{Kantorovich lifting} of the functor $F$ is the functor $\LiftedFunctor{F}\colon \PMet\to\PMet$ defined as $\LiftedFunctor{F}(X,d) = (FX,\Kantorovich{F}{d})$ and $\LiftedFunctor{F}f = Ff$. 
\end{savetheorem}

\noindent It is easy to show that $\Kantorovich{F}{d}$ is indeed a
pseudometric. Since $\LiftedFunctor{F}$ inherits the preservation of
identities and composition of morphisms from $F$ we can prove that
nonexpansive functions are mapped to nonexpansive functions and
isometries to isometries.

\begin{savetheorem}{proposition}{prop:kantorovich-lifting-preserves-isometries}
The Kantorovich lifting $\LiftedFunctor{F}$ of a functor $F$ preserves isometries. 
\end{savetheorem}

\noindent We chose the name \emph{Kantorovich} because our definition is reminiscent of the Kantorovich pseudometric in probability theory. If we take the proper combination of functor and evaluation function, we can recover that pseudometric (in the discrete case) as the first instance for our framework. 

\begin{savetheorem}[Probability Distribution Functor]{example}{exa:probability-distribution-functor}
We take $\top = 1$ and the probability distribution functor $\Distributions $ from \exampleref{ex:probabilistic-1}. As evaluation function we take the expected value of the identity on $[0,1]$, i.e. for any $P \in \Distributions [0,1]$ we have $\ev_\Distributions (P) = \ExpectedValue{P}{\id_{[0,1]}} = \sum_{x\in [0,1]} x\cdot P(x)$ yielding $\EvaluationFunctor{\Distributions }g (P) = \ExpectedValue{P}{g} = \sum_{x\in [0,1]} g(x)\cdot P(x)$ for all $g\colon X \to [0,1]$. For every pseudometric space $(X,d)$ we obtain the discrete Kantorovich pseudometric $\Kantorovich{\Distributions }{d}\colon \Distributions X \times \Distributions X \to [0,1]$, defined as, for all $P_1, P_2 \in \Distributions X$, $\Kantorovich{\Distributions }{d}(P_1,P_2) = \sup\{ \sum_{x \in X} f(x)\cdot \big(P_1(x) - P_2(x)\big) \mid f \colon (X,d) \nonexpansiveTo ([0,1],d_e)\}$.
\end{savetheorem}

\noindent In general Kantorovich liftings do not preserve metrics, as shown by the following example.

\begin{savetheorem}{example}{exa:kantorovich-lifting-does-not-preserve-metrics}
Let $F\colon \Set \to \Set$ be given as $FX = X \times X$ on sets and $Ff = f \times f$ on functions and take $\top = \infty$, $\ev_F\colon F[0,\infty] \to [0,\infty], \ev_F(r_1, r_2) = r_1+r_2$. For a metric space $(X,d)$ with $|X|\geq 2$ let $t_1 = (x_1, x_2) \in FX$ with $x_1 \not = x_2$ and define $t_2 := (x_2, x_1)$. Clearly $t_1 \not = t_2$ but for every nonexpansive function $f\colon (X,d) \nonexpansiveTo (\reals, d_e)$ we have $\EvaluationFunctor{F}f(t_1) = f(x_1) + f(x_2) = f(x_2) + f(x_1) = \EvaluationFunctor{F}f(t_2)$ and thus $\Kantorovich{F}{d}(t_1,t_2) = 0$.
\end{savetheorem}

\section{Wasserstein Pseudometric and Kantorovich-Rubinstein Duality}
\label{sec:wasserstein}
We have seen that our first lifting approach bears close resemblance to the original Kantorovich pseudometric on probability measures. In that context there exists another pseudometric, the Wasserstein pseudometric, which under certain conditions coincides with the Kantorovich pseudometric. We will define a generalized version of the Wasserstein pseudometric and compare it with our generalized Kantorovich pseudometric. To do that we first need to define how we can couple elements of $FX$.

\begin{definition}[Coupling]
\label{def:coupling}
Let $F \colon \Set \to \Set$ be a functor and $n \in \N$. Given a set $X$ and $t_i \in FX$ for $1 \leq i \leq n$ we call an element $t \in F(X^n)$ such that $F\pi_i(t) = t_i$ a \emph{coupling} of the $t_i$ (with respect to $F$). We write $\Couplings{F}(t_1, t_2, \dots, t_n)$ for the set of all these couplings. 
\end{definition}

\noindent If $F$ preserves weak pullbacks, we can define new couplings based on given ones.

\begin{savetheorem}[Gluing Lemma]{lemma}{lem:coupling}
Let $F \colon \Set \to \Set$ be a weak pullback preserving functor, $X$ a set, $t_1, t_2, t_3 \in FX$, $t_{12} \in \Couplings{F}(t_1,t_2)$, and $t_{23} \in \Couplings{F}(t_2,t_3)$ be couplings. Then there is a coupling $t_{123} \in \Couplings{F}(t_1, t_2, t_3)$ such that $F(\langle \pi_1^3, \pi_2^3\rangle)(t_{123}) = t_{12}$ and $F(\langle\pi_2^3, \pi_3^3\rangle)(t_{123}) = t_{23}$.
\end{savetheorem}

\noindent This lemma already hints at the fact that our new lifting will only work for weak pullback preserving functors, which is a standard requirement in coalgebra. In addition to that we have to impose three extra conditions on the evaluation functions.

\begin{savetheorem}[Well-Behaved Evaluation Function]{definition}{def:well-behaved}
Let $\ev_F$ be an evaluation function for a functor $F\colon \Set \to \Set$. We call $\ev_F$ \emph{well-behaved} if it satisfies the following conditions:
\begin{enumerate}
\item $\EvaluationFunctor{F}$ is monotone, i.e., for $f,g\colon X\to\reals$ with 
$f\le g$, we have $\EvaluationFunctor{F}f\le\EvaluationFunctor{F}g$.
\item For each $t\in F(\reals^2)$ it holds that $d_e(\ev_F(t_1), \ev_F(t_2)) \leq \EvaluationFunctor{F}d_e(t)$ for $t_i : = F\pi_i(t)$.
\item $\ev_F^{-1}[\set{0}] = Fi[F\{0\}]$ where $i \colon \set{0} \hookrightarrow\reals$ is the inclusion map. 
\end{enumerate}
\end{savetheorem}

\noindent While the first condition of this definition is quite natural, the other two need to be explained. Condition~2 is needed to ensure that $\EvaluationFunctor{F}\id_\reals = \ev_F\colon F\reals\to\reals$ is nonexpansive once $d_e$ is lifted to $F\reals$ (cf.\ the intuition behind the Kantorovich lifting, where we ensure that $\EvaluationFunctor{F}f$ is nonexpansive whenever $f$ is nonexpansive). Furthermore 
Condition~3 intuitively says that exactly the elements of $F\{0\}$ are mapped to $0$ via $\ev_F$. Before we define the Wasserstein pseudometric and the corresponding lifting, we take a look at an example of a functor together with a well-behaved evaluation function.

\begin{savetheorem}[Finite Powerset Functor]{example}{ex:evaluation-max}
Let $\top = \infty$. We take the finite powerset functor $\PowersetFinite$ with evaluation function $\max\colon \PowersetFinite(\prealinf) \to \prealinf$ with $\max \emptyset = 0$. This evaluation function is well-behaved whereas $\min\colon \PowersetFinite(\prealinf) \to \prealinf$ is not well-behaved.
\end{savetheorem}

\begin{definition}[Wasserstein Pseudometric \& Wasserstein Lifting]
\label{def:wasserstein}
Let $F\colon \Set \to \Set$ be a weak-pullback preserving functor with well-behaved evaluation function $\ev_F$. For every pseudometric space $(X,d)$ the \emph{Wasserstein pseudometric} on $FX$ is the function $\Wasserstein{F}{d} \colon FX \times FX \to \reals$ given by, for all $t_1,t_2 \in FX$,  
\begin{align*}
	\Wasserstein{F}{d}(t_1, t_2) := \inf \{\EvaluationFunctor{F}d(t) \mid t \in \Couplings{F}(t_1,t_2)\}\,.
\end{align*}
We define the \emph{Wasserstein lifting} of $F$ to be the functor $\LiftedFunctor{F}\colon \PMet\to\PMet$, $\LiftedFunctor{F}(X,d) = (FX,\Wasserstein{F}{d})$, $\LiftedFunctor{F}f = Ff$.
\end{definition}

\noindent This time it is not straightforward to prove that $\Wasserstein{F}{d}$ is a pseudometric, so we explicitly provide the following result. Its proof relies on all properties of well-behavedness of $\ev_F$ and uses \lemmaref{lem:coupling} which explains why we need a weak pullback preserving functor.

\begin{savetheorem}{proposition}{prop:wasserstein-is-pseudometric}
The Wasserstein pseudometric is a well-defined pseudometric on $FX$.
\end{savetheorem}

\noindent In contrast to that, it is not hard to show functoriality of $\LiftedFunctor{F}$ and, as in the Kantorovich case, the lifted functor preserves isometries.

\begin{savetheorem}{proposition}{prop:wasserstein-preserves-isometries}
The Wasserstein lifting $\LiftedFunctor{F}$ of a functor $F$ preserves isometries. 
\end{savetheorem}

\noindent In contrast to our previous approach, metrics are preserved in certain situations.

\begin{savetheorem}[Preservation of Metrics]{proposition}{prop:preservation}
Let $(X,d)$ be a metric space and $F$ be a functor. If the infimum in \definitionref{def:wasserstein} is a minimum for all $t_1,t_2 \in FX$ where $\Wasserstein{F}{d}(t_1,t_2) = 0$ then $\Wasserstein{F}{d}$ is a metric, thus also $\LiftedFunctor{F}(X,d) = (FX,\Wasserstein{F}{d})$ is a metric space.
\end{savetheorem}

\noindent Please note that a similar restriction for the Kantorovich lifting (i.e. requiring that the supremum in \definitionref{def:kantorovich} is a maximum) does not yield preservation of metrics: In \exampleref{exa:kantorovich-lifting-does-not-preserve-metrics} the supremum is always a maximum but we do not get a metric.

Let us now compare both lifting approaches. Whenever it is defined, the Wasserstein pseudometric is an upper bound for the Kantorovich pseudometric.

\begin{savetheorem}{proposition}{prop:wasserstein-vs-kantorovich}
Let  $F$  be a  weak pullback preserving functor with well-behaved evaluation function. Then for all pseudometric spaces $(X,d)$ it holds that $\Kantorovich{F}{d} \leq \Wasserstein{F}{d}$.
\end{savetheorem}

\noindent In general this inequality may be strict in general, as the following example shows.

\begin{savetheorem}{example}{exa:kantorovich-lifting-does-not-preserve-metrics2}
The functor of
\exampleref{exa:kantorovich-lifting-does-not-preserve-metrics}
preserves weak pullbacks and the evaluation function is
well-behaved. We continue the example and take $t_1=(x_1,x_2)$,
$t_2=(x_2,x_1)$. The unique coupling $t \in \Gamma_F(t_1,t_2)$ is
$t=(x_1, x_2, x_2, x_1)$. Using that $d$ is a metric we conclude that
$\Wasserstein{F}{d}(t_1, t_2) = \EvaluationFunctor{F}d(t) = d(x_1,x_2)
+ d(x_2,x_1) = 2d(x_1,x_2) > 0 = \Kantorovich{F}{d}(t_1,t_2)$.
\end{savetheorem}

\noindent When the inequality can be replaced by an equality we will
in the following say that the Kantorovich-Rubinstein duality holds.
In this case we obtain a canonical notion of distance on $FX$, given a
pseudometric space $(X,d)$. To calculate the distance of $t_1,t_2\in
FX$ it is then enough to find a nonexpansive function $f\colon
(X,d)\nonexpansiveTo (\reals,d_e)$ and a coupling $t \in
\Gamma_F(t_1,t_2)$ such that
$d_e(\EvaluationFunctor{F}f(t_1),\EvaluationFunctor{F}f(t_2)) =
\EvaluationFunctor{F}d_e(t)$. Then, due to
\propositionref{prop:wasserstein-vs-kantorovich}, this value equals
$\Kantorovich{F}{d}(t_1,t_2) = \Wasserstein{F}{d}(t_1,t_2)$.  We will
now take a look at some examples where the duality holds.

\begin{example}[Identity Functor] 
\label{ex:kant-rubinst}
Take $F = \Id$ with the identity evaluation map $\ev_\Id = \id_\reals$. For any $t_1,t_2\in X$, $t := (t_1, t_2)$ is the unique coupling of $t_1, t_2$. Hence, $\Wasserstein{F}{d}(t_1, t_2) = d(t_1,t_2)$. With the function $d(t_1,\_) \colon (X,d) \nonexpansiveTo (\reals,d_e)$ we obtain duality because we have $d(t_1,t_2) = d_e(d(t_1, t_1), d(t_1,t_2)) \leq \Kantorovich{F}{d}(t_1,t_2) \leq \Wasserstein{F}{d}(t_1,t_2) = d(t_1,t_2)$ and thus equality. Similarly, if we define $\ev_\Id(r) = c\cdot r$ for $r\in \reals$, $0<c\leq1$, the Kantorovich and Wasserstein liftings coincide and we obtain the discounted distance $\Kantorovich{F}{d}(t_1,t_2) = \Wasserstein{F}{d}(t_1,t_2) = c\cdot d(t_1,t_2)$.
\end{example}

\begin{savetheorem}[Probability Distribution Functor]{example}{exa:probability-distribution-functor2}
The functor $\Distributions $ of \exampleref{exa:probability-distribution-functor} preserves weak pullbacks \cite{Sok11} and the evaluation function $\ev_\Distributions $ is well-behaved. We recover the usual Wasserstein pseudometric 
$\Wasserstein{\Distributions }{d}(P_1,P_2)= \inf \{\sum_{x_1,x_2 \in X} d(x_1,x_2) \cdot P(x_1, x_2) \mid P \in \Couplings{\Distributions }(P_1, P_2)\}$
and the Kantorovich-Rubinstein duality \cite{v:optimal-transport} from transportation theory for the discrete case.
\end{savetheorem}

\begin{savetheorem}[Finite Powerset Functor \& Hausdorff Pseudometric]{example}{exa:hausdorff}
Let $\top = \infty$, $F = \PowersetFinite$ with evaluation map $\ev_{\PowersetFinite}\colon \PowersetFinite(\prealinf) \to \prealinf$, $\ev_{\PowersetFinite}(R) = \max R$ with $\max \emptyset = 0$ (as in \exampleref{ex:evaluation-max}). In this setting we obtain duality and both pseudometrics are equal to the Hausdorff pseudometric $d_H$ on $\PowersetFinite(X)$ which is defined as, for all $X_1,X_2\in \PowersetFinite(X)$,
\begin{align*}
	 d_H(X_1,X_2) = \max\left\{\max_{x_1\in X_1} \min_{x_2\in X_2} d(x_1,x_2), \max_{x_2\in X_2} \min_{x_1\in X_1} d(x_1,x_2) \right\}\,.
\end{align*}
Note that the distance is $\infty$, if either $X_1$ or $X_2$ is empty.
\end{savetheorem}

\noindent It is also illustrative to consider the countable powerset
functor.  Using the supremum as evaluation function, one obtains again
the Hausdorff pseudometric (with supremum/infimum replacing
maximum/minimum). However,  in this case the Hausdorff
distance of different countable sets might be $0$, even if we lift
a metric. This shows that in general the Wasserstein lifting does not preserve metrics but we need an extra condition, e.g. the one in \propositionref{prop:preservation}.

\section{Lifting Multifunctors}
\label{sec:multifunctors}
Our two approaches can easily be generalized\footnote{The details are spelled out in \appendixref{app:multifunctors}, here we provide just the basic ideas.} to lift a multifunctor $F\colon\Set^n\to \Set$ (for $n \in \N$) in a similar sense as given by \definitionref{def:lifting} to a multifunctor $\LiftedFunctor{F}\colon \PMet^n \to \Set$. The only difference is that we start with $n$ pseudometric spaces instead of one. Now we need an \emph{evaluation function} $\ev_F\colon F(\reals, \dots, \reals) \to \reals$ which we call \emph{well-behaved} if it satisfies conditions similar to \definitionref{def:well-behaved} and which gives rise to an evaluation multifunctor $\EvaluationFunctor{F} \colon(\Set/\reals)^n \to \Set/\reals$. Given $t_1,t_2 \in F(X_1,\dots,X_n)$ we write again $\Couplings{F}(t_1,t_2) \subseteq F(X_1^2,\dots, X_n^2)$ for the set of couplings which is defined analogously to \definitionref{def:coupling}. For pseudometrics $d_i\colon X_i^2\to \reals$, we can then define the \emph{Kantorovich/Wasserstein pseudometric} $\KantorovichMulti{F}{d}{1,\dots,n}, \WassersteinMulti{F}{d}{1,\dots,n}\colon F(X_1,\dots,X_n)\times F(X_1,\dots,X_n)\to \reals$, as $\KantorovichMulti{F}{d}{1,\dots,n}(t_1,t_2) := \sup \{d_e(\EvaluationFunctor{F}(f_1,\dots,f_n)(t_1),\EvaluationFunctor{F}(f_1,\dots,f_n)(t_2)) \mid {f_i\colon (X_i,d_i) \nonexpansiveTo (\reals,d_e)}\}$ and $\WassersteinMulti{F}{d}{1,\dots,n}(t_1, t_2) := \inf \{\EvaluationFunctor{F}(d_1,\dots,d_n)(t) \mid t \in \Couplings{F}(t_1,t_2)\}$. This setting grants us access to new examples such as the product and the coproduct bifunctors.

\begin{savetheorem}[Product Bifunctor]{example}{exa:product-bifunctor}
For the product bifunctor $F\colon \Set^2 \to \Set$ where $F(X_1,X_2) = X_1\times X_2$ and $F(f_1,f_2) = f_1 \times f_2$ we consider the evaluation function $\max \colon\reals^2\to\reals$ and for fixed parameters $c_1,c_2 \in (0,1]$ and $p \in \mathbb{N}$ the function $\rho\colon\reals^2\to\reals$, $\rho(x_1,x_2) = (c_1 x_1^p + c_2 x_2^p)^{1/p}$. These functions are well-behaved, the Kantorovich-Rubinstein duality holds and the supremum [infimum] of the Kantorovich [Wasserstein] pseudometrics is always a maximum [minimum]. For the first function we obtain the $\infty$-product pseudometric $d_\infty((x_1,x_2),(y_1,y_2)) = \max(d_1(x_1,y_1), d_2(x_2,y_2))$ and for the other function the weighted $p$-product pseudometric $d_p((x_1,x_2),(y_1,y_2)) = (c_1d_1^p(x_1,y_1)+c_2d_2^p(x_2,y_2))^{1/p}$.
\end{savetheorem}

\noindent Note that the pseudometric space $(X_1\times X_2,d_\infty)$ is the usual binary (category theoretic) product of $(X_1,d_1)$ and $(X_2, d_2)$. Similarly, we can also obtain the binary coproduct.

\begin{savetheorem}[Coproduct Bifunctor]{example}{exa:coproduct-bifunctor}
For the coproduct bifunctor $F\colon \Set^2 \to \Set$, where $F(X_1,X_2) = X_1 + X_2 = X_1\times\{1\} \cup X_2\times \{2\}$ and $F(f_1,f_2) = f_1+f_2$ we take the evaluation function $\ev_F\colon\reals+\reals\to\reals$, $\ev_F(x,i) = x$. This function is well-behaved, the Kantorovich-Rubinstein duality holds and the supremum of the Kantorovich pseudometric is always a maximum whereas the infimum of the Wasserstein pseudometric is a minimum if and only if any coupling of the two elements exists. We obtain the coproduct pseudometric $d_+$ where $d_+((x_1,i_1), (x_2, i_2))$ is equal to $d_{i}(x_1,x_2)$ if $i_1=i_2=i$ and equal to $\top$ otherwise.
\end{savetheorem}

\section{Final Coalgebra \& Coalgebraic Behavioral Pseudometrics}
\label{sec:final-coalgebra}

In this section we assume an arbitrary lifting $\LiftedFunctor{F}\colon \PMet \to \PMet$ of an endofunctor $F$ on $\Set$. For any pseudometric space $(X,d)$ we write
$\LiftedMetric{F}{d}$ for the pseudometric obtained by applying
$\LiftedFunctor{F}$ to $(X,d)$. Such a lifting can be obtained as
described earlier, but also by taking a lifted multifunctor and
fixing all parameters apart from one, or by the composition of such
functors. The following result ensures that if $\kappa \colon \Omega \to F
\Omega$ is a final $F$-coalgebra, then there is also a final
$\LiftedFunctor{F}$-coalgebra which is constructed by simply enriching
$\Omega$ with a pseudometric $d_\Omega$. 

\begin{savetheorem}{theorem}{thm:final-coalgebra}
Let $\LiftedFunctor{F}\colon\PMet\to\PMet$ be a lifting of a functor $F\colon \Set\to \Set$ which has a final coalgebra $\kappa\colon\Omega\to F\Omega$. For every ordinal $i$ we construct a pseudometric $d_i\colon\Omega\times\Omega\to\reals$ as follows: $d_0 := 0$ is the zero pseudometric, $d_{i+1} := \LiftedMetric{F}{d_i}\circ(\kappa\times\kappa)$ for all ordinals $i$ and $d_j = \sup_{i<j} d_i$ for all limit ordinals $j$.
This sequence converges for some ordinal $\theta$, i.e $d_\theta = \LiftedMetric{F}{d_\theta}\circ(\kappa\times\kappa)$. Moreover $\kappa\colon (\Omega,d_\theta) \nonexpansiveTo (F\Omega,\LiftedMetric{F}{d_\theta})$ is the final $\LiftedFunctor{F}$-coalgebra.
\end{savetheorem}

\noindent We noted that for any set $X$, the set of pseudometrics over
$X$, with pointwise order, is a complete lattice. Moreover the lifting
$\LiftedFunctor{F}$ induces a monotone function $\_^F$ which maps any
pseudometric $d$ on $X$ to $d^F$ on $FX$. If, additionally, such function is
$\omega$-continuous, i.e., it preserves the supremum of
$\omega$-chains, the construction in \theoremref{thm:final-coalgebra}
will converge in at most $\omega$ steps, i.e., $d_\theta=d_\omega$.
We show in \apxpropositionref{prop:w-continuity} that the liftings induced by the
finite powerset functor and the probability distribution functor with
finite support are $\omega$-continuous. The arguments used for
convergence here suggests a connection with the work
in~\cite{vBHMW:RDMWC}, which provides fixed-point results for metric
functors which are not locally contractive.

Beyond equivalences of states, in $\PMet$ we can measure the distance
of behaviors in the final coalgebra. More precisely, the
\emph{behavioral distance} of two states $x,y\in X$ of some coalgebra
$\alpha\colon X\to FX$ is defined via the pseudometric
$\mathit{bd}(x,y) = d_\theta(\final{x},\final{y})$. Such distances can
be computed analogously to $d_\theta$ above, replacing $\kappa\colon
\Omega \to F \Omega$ by $\alpha$. This way we do not need to explore the entire final coalgebra (which might be too large) but can restrict to the interesting part.

\begin{savetheorem}{theorem}{thm:comp-dist}
Let the chain of the $d_i$ converge in $\theta$ steps and
$\overline{F}$ preserve isometries. Let furthermore $\alpha\colon X\to
FX$ be an arbitrary coalgebra. For all ordinals $i$ we define a pseudometric $e_i\colon X\times X\to \reals$ as follows: $e_0$ is the zero pseudometric, $e_{i+1} = e_i^F\circ (\alpha\times\alpha)$ for all ordinals $i$ and $e_j = \sup_{i<j} e_i$ for all limit ordinals $j$. Then we reach a fixed point after $\zeta \leq \theta$ steps, i.e. $e_\zeta = e_\zeta^F\circ (\alpha\times\alpha)$, such that $\mathit{bd} = e_\zeta$.
\end{savetheorem}

\noindent Since $d_\theta$ is a pseudometric, we have that if
$\final{x}=\final{y}$ then $bd(x,y)=0$. The other direction does
not hold in general: for this $d_\theta$ has to be a proper metric.
\theoremref{thm:d-omega-is-metric} at the end of this section
provides sufficient conditions guaranteeing this property.

To this aim, we proceed by recalling the final coalgebra construction
via the final chain which was first presented in the dual setting
(free/initial algebra).
\begin{definition}[Final Coalgebra Construction \cite{Ada74}]
Let $\mathbf{C}$ be a category with terminal object $\one$ and limits of ordinal-indexed cochains. For any functor $F\colon \mathbf{C} \to \mathbf{C}$ the \emph{final chain} consists of objects $W_i$ for all ordinals $i$ and \emph{connection morphisms} $p_{i,j} \colon W_j \to W_i$ for all ordinals $i \le j$. The objects are defined as $W_0 := \one$, $W_{i+1} := FW_i$ for all ordinals $i$, and $W_j := \lim_{i<j}W_i$ for all limit ordinals $j$. The morphisms are determined by $p_{0,i} :=\ !\colon W_i \to \one$, $p_{i,i} = \id_{W_i}$ for all ordinals $i$, $p_{i+1,j+1} := Fp_{i,j}$ for all ordinals $i < j$ and if $j$ is a limit ordinal the $p_{i,j}$ are the morphisms of the limit cone. They satisfy $p_{i,k} = p_{i,j} \circ p_{j,k}$ for all ordinals $i \leq j \leq k$. We say that the chain \emph{converges} in $\lambda$ steps if $p_{\lambda,\lambda+1}\colon W_{\lambda+1} \to W_\lambda$ is an iso. 
\end{definition}

\noindent This construction does not necessarily converge, but if it does, we get a final coalgebra. 

\begin{proposition}[\cite{Ada74}] 
Let $\mathbf{C}$ be a category with terminal object $\one$ and limits of ordinal-indexed cochains. If the final chain of a functor $F\colon \mathbf{C} \to \mathbf{C}$ converges in $\lambda$ steps then $p_{\lambda, \lambda+1}^{-1}\colon W_\lambda \to FW_\lambda$ is the final coalgebra.
\end{proposition}

\noindent We now show under which circumstances $d_\theta$ is a metric and how our construction relates to the construction of the final chain.

\begin{savetheorem}{theorem}{thm:d-omega-is-metric}
Let $\LiftedFunctor{F}\colon\PMet\to\PMet$ be a lifting of a functor $F\colon \Set\to \Set$ which has a final coalgebra $\kappa\colon\Omega\to F\Omega$. Assume that $\LiftedFunctor{F}$ preserves isometries and metrics, that the final chain for $F$ converges and the chain of the $d_i$ converges in $\theta$ steps. Then $d_\theta$ is a metric, i.e. for $x,y\in \Omega$ we have $d_\theta(x,y) = 0\iff x=y$.
\end{savetheorem}
 
\noindent We will now get back to the examples studied at the beginning of the paper (\exampleref{ex:probabilistic-1} and \exampleref{ex:metric-ts-1}) and discuss in which sense they are instances of our framework.

\begin{example}[Probabilistic Transition System, revisited]
  \label{ex:probabilistic-2}
  To model the behavioral distance from
  \exampleref{ex:probabilistic-1} in our framework, we set $\top=1$
  and proceed to lift the following three functors: we first consider
  the identity functor $\Id$ with evaluation map $\ev_\Id\colon
  [0,1]\to [0,1]$, $\ev_\Id(z) = c\cdot z$ in order to integrate the
  discount (\exampleref{ex:kant-rubinst}). Then, we take the
  coproduct with the singleton metric space
  (\exampleref{exa:coproduct-bifunctor}).  The combination of the two
  functors yields the discrete version of the refusal functor of
  \cite{vBW06}, namely $\LiftedFunctor{R}(X,d) = (X+\mathbf{1},
  \hat{d})$ where $\hat{d}$ is taken from
  \exampleref{ex:probabilistic-1}.  Finally, we lift the probability
  distribution functor $\Distributions $ to obtain
  $\LiftedFunctor{\Distributions }$
  (\exampleref{exa:probability-distribution-functor}). All functors
  satisfy the Kantorovich-Rubinstein duality and preserve metrics.

It is readily seen that $\LiftedFunctor{\Distributions}(\LiftedFunctor{R}(X,d))=(\Distributions(X+1), \overline{d})$, where $\overline{d}$ is defined as in \exampleref{ex:probabilistic-1}).  Then, the least solution of $d(x,y) = \overline{d}(\alpha(x),\alpha(y))$ can be computed as in \theoremref{thm:comp-dist}.
\end{example}

\begin{example}[Metric Transition Systems, revisited]
\label{ex:metric-ts-2}
To obtain propositional distances in metric transition systems we set $\top = \infty$. We also define, for the auxiliary functor $G$, an evaluation function $\ev_G\colon G(\prealinf,\dots,\prealinf)\to\prealinf$ with $\ev_G(u) = \max_{r\in\Sigma} u(r)$. Let $\LiftedFunctor{G}$ be the corresponding lifted functor. It can be shown, similarly to \exampleref{exa:product-bifunctor}, that the Kantorovich-Rubinstein duality holds and metrics are preserved. We instantiate the given pseudometric spaces $(M_{r_i},d_{r_i})$ as parameters and obtain the functor $\LiftedFunctor{F}(X,d) = \LiftedFunctor{G}((M_{r_1},d_{r_1}),\dots,(M_{r_n},d_{r_n})) \times \LiftedFunctor{\PowersetFinite}(X,d)$ (for the lifting of the powerset functor see \exampleref{exa:hausdorff}).  Then, via \theoremref{thm:comp-dist}, we obtain exactly the least solution of \eqref{eq:Hausd} in \exampleref{ex:metric-ts-1}.
\end{example}

\section{Related and Future Work}
The ideas for our framework are heavily influenced by work on quantitative variants of (bisimulation) equivalence of probabilistic systems. In that context at first Giacalone et al. \cite{GJS90} observed that probabilistic bisimulation \cite{LS89} is too strong and therefore introduced a metric based on the notion of $\epsilon$-bisimulations.

Using a logical characterization of bisimulation for labelled Markov processes (LMP) \cite{DEP02}, Desharnais et al. defined a family of metrics between these LMPs \cite{DGJP04} via functional expressions: if evaluated on a state of an LMP, such a functional expression measures the extent to which a formula is satisfied in that state.  A different, coalgebraic approach, which inspired ours, is used by van Breugel et al. \cite{vBW06}. As presented in more detail in the examples above, they define a pseudometric on probabilistic systems via the Kantorovich pseudometric for probability measures. Moreover, they show in \cite{vBW05} that this metric is related to the logical pseudometric by Desharnais et al.

Our framework provides a toolbox to determine behavioral distances for different types of transition systems modeled as coalgebras.
Moreover, the liftings introduced in this paper pave the way to extend several coalgebraic methods to reason about quantitative properties of systems.
For instance the bisimulation proof principle, which allows to check
behavioral equivalence, assumes a specific meaning in $\PMet$: every
coalgebra $\alpha \colon (X,d) \to \LiftedFunctor{F}(X,d)$
\emph{coinductively proves} that the behavioral distance $bd$ of the
underlying $F$-coalgebra on $\Set$ is smaller or equal than $d$.
Indeed, since $\finalO$ is nonexpansive, $d \geq
d_\theta(\finalO,\finalO)=bd$. This principle, which has already
been stated in different formulations (see
e.g.~\cite{DengCPP06,Desharnais02themetric,BreugelSW08}), can now be
enhanced via \emph{up-to techniques} by exploiting the liftings
introduced in this paper and the coalgebraic understanding of such
enhancements given in~\cite{DBLP:journals/corr/BonchiPPR14}.

Since up-to techniques can exponentially improve algorithms for
equivalence-checking, we hope that they could also optimize some of
the algorithms for computing (or approximating) behavioral
distances~\cite{vBW06,BreugelSW08,DBLP:conf/fossacs/ChenBW12,DBLP:conf/tacas/BacciBLM13}.
At this point, it is worth recalling that the Kantorovich-Rubinstein
duality has been exploited in~\cite{vBW06} for defining one of these
algorithms: the characterization given by the Wasserstein metric
allows to reduce to linear programming.

Another line of research potentially stemming from our work concerns the
so-called \emph{abstract GSOS}~\cite{DBLP:journals/tcs/Klin11} which
provides abstract coalgebraic conditions ensuring compositionality of
behavioral equivalence (with respect to some operators). By taking our lifting to
$\PMet$, abstract GSOS guarantees the nonexpansiveness of behavioral
distance, a property that has captured the interest of several
researchers~\cite{DGJP04,DBLP:journals/corr/GeblerT13}.  The main
technical challenge would be to lift to $\PMet$ not only functors, but
also distributive laws. Lifting of distributive laws would also be
needed for defining \emph{linear behavioral distances}, exploiting the
coalgebraic account of trace semantics based on Kleisli
categories~\cite{HJS07}.

We finally observe that the chains of Theorems~\ref{thm:final-coalgebra} and~\ref{thm:comp-dist} can be understood in terms of fibrations along the lines of~\cite{HasuoCKJ13}. A detailed comparison with~\cite{HasuoCKJ13} can be found in \appendixref{app:fibrations}.

\subparagraph*{Acknowledgements}
The authors are grateful to Franck van Breugel, Neil Ghani and  Daniela Petri{\c s}an for several precious suggestions and inspiring discussions.
The second author acknowledges the support by project \texttt{ANR 12IS0} \texttt{2001 PACE}.

\bibliography{paper-129}

%% file: apx-fibrations.tex
%!TEX root = paper-129-extended.tex
\section{A Fibrational Perspective}\label{app:fibrations}
A recent work~\cite{HasuoCKJ13} studies final chains in fibrations. This is interesting for our work, since the forgetful functor $U \colon \PMet \to \Set$ is a poset fibration: the fibre above a set $X$ (denoted by $\PMet_X$) is the poset of all pseudometric spaces over $X$ with the order $d_1\sqsubseteq d_2$ iff $d_2\leq d_1$; for a function $f\colon X \to Y$, the reindexing functor $f^*\colon \PMet_Y \to \PMet_X$ maps a space $(Y,d)$ into $(X,d^*)$ where $d^*(x_1,x_2)=d(f(x_1),f(x_2))$, namely the cartesian lifting $\tilde{f_{d}}\colon (X,d^*) \to (Y,d)$ is an isometry. By virtue of Lemma 3.5~in~\cite{HasuoCKJ13}, one can readily check that this fibration has fiberwise limits. Indeed, $\PMet$ is complete and $U$ preserves limits since it has a left adjoint (mapping a set $X$ into the space $(X,d^{\top})$ where $d^{\top}(x_1,x_2)=\top$ for all $x_1\neq x_2$).

%Since both $\Set$ and $\PMet$ are complete (see Proposition \ref{prop:PMetcomplete}) and $U$ preserve limits 
%As shown in Proposition \ref{prop:PMetcomplete},  Therefore, by virtue of Lemma 3.5 in \cite{}, The fibration $U$ has fiberwise limits.

In this setting, a pair of functors $F\colon \Set\to\Set$, $\LiftedFunctor{F}\colon \PMet\to\PMet$ is a map of fibrations iff $\LiftedFunctor{F}$ is a lifting of $F$ and, additionally, $\LiftedFunctor{F}$ preserves isometries, a property enjoyed by both Kantorovich and Wasserstein liftings (Propositions~\ref{prop:kantorovich-lifting-preserves-isometries} and~\ref{prop:wasserstein-preserves-isometries}). By Proposition~4.1 in~\cite{HasuoCKJ13}, 
$U$ lifts to a fibration $\LiftedFunctor{U}\colon
\Coalg{\LiftedFunctor{F}}\to \Coalg{F}$ from
$\LiftedFunctor{F}$-coalgebras to $F$-coalgebras and, again by virtue
of Lemma 3.5 in~\cite{HasuoCKJ13}, the final object of $\Coalg{\LiftedFunctor{F}}$ is the final object of $\Coalg{\LiftedFunctor{F}}_{\kappa}$, the fibre  above the final $F$-coalgebra $\kappa\colon \Omega\to F\Omega$.

For any coalgebra $\alpha\colon X\to FX$, the fibre above $\alpha$ is isomorphic to the category $\Coalg{\alpha^*\circ \LiftedFunctor{F}}$ of coalgebras for the endofunctor $\alpha^*\circ \LiftedFunctor{F}\colon \PMet_X\to \PMet_X$ (Proposition 4.2 in~\cite{HasuoCKJ13}). Therefore the final $\LiftedFunctor{F}$-coalgebra is just the final coalgebra for the endofunctor $\kappa^*\circ \LiftedFunctor{F}\colon \PMet_{\Omega}\to \PMet_{\Omega}$. By unfolding the definitions, one can see that the sequence of $d_i$ in Theorem~\ref{thm:final-coalgebra} is indeed the final chain for $\kappa^*\circ \LiftedFunctor{F}$. Similarly, the sequence of $e_i$ in Theorem~\ref{thm:comp-dist} is just the final chain for $\alpha^*\circ \LiftedFunctor{F}$.

Theorem~3.7 in~\cite{HasuoCKJ13} provides sufficient conditions for a fibration to ensure the convergence of these chains in $\omega$~steps. Unfortunately, $U$ does not fulfill this conditions, as it is not ``well-founded''.

%% file: apx-proofs.tex
%!TEX root = paper-129-extended.tex
\setcounter{section}{15} % P for Proof ;)
\section{Proofs}
\label{sec:proofs}
Here we provide proofs for the soundness of our definitions (where needed), the stated theorems, propositions, lemmas, examples and also for all claims made in the in-between texts. If a theorem environment starts with \textcolor{orange}{$\blacksquare$} (orange square) it has been stated in the main text and is repeated here for convenience of the reader (using the numbering from the main text). Otherwise it is a new statement which clarifies/justifies claims made in the main text and its number starts with \emph{P}.

\stepcounter{subsection}
\renewcommand{\sectionautorefname}{Appendix}
\renewcommand{\subsectionautorefname}{Appendix}

\subsection{\nameref{sec:preliminaries}}
\label{apx:preliminaries}
In the following lemma we rephrase the well-known fact that for $a,b,c \in [0,\infty)$ we have $|a-b| \leq c \iff a-b \leq c\, \land\, b-a \leq c$ to include the cases where $a,b,c$ might be $\infty$.

\begin{lemma_apx}
\label{lem:sum-vs-dist}
For $a,b,c\in\prealinf$ we have $d_e(a,b)\leq c \iff (a \leq b+c)\, \land\, (b \leq a+c) $.
\end{lemma_apx}

\begin{proof}
This equivalence is obvious for $a,b,c\in [0, \infty)$. In the cases $a,b \in \prealinf$, $c=\infty$ or $a=b=\infty$, $c \in \prealinf$ both sides of the equivalence are true whereas for the cases $a= \infty$, $b,c \in [0,\infty)$ or $b= \infty$, $a,c \in [0,\infty)$ both sides are false.
\end{proof}

\restate{lem:alt-char-triangle}
\begin{proof}
We show both implications for all $x,y,z \in X$.\\
($\Rightarrow$) Using the triangle equation and symmetry we know that $d(x,y) \leq d(x,z) + d(y,z)$ and $d(x,z) \leq d(x,y) + d(y,z)$. With \lemmaref{lem:sum-vs-dist} we conclude that $d_e(d(x,y),d(x,z)) \leq d(y,z)$.

\noindent ($\Leftarrow$) Using $d(x,x) = 0$, the triangle inequality for $d_e$ and nonexpansiveness of $d(x,\_)$ we get $d(x,z) = d_e(d(x,x),d(x,z)) \leq d_e(d(x,x),d(x,y)) + d_e(d(x,y),d(x,z)) \leq d(x,y) + d(y,z)$. 
\end{proof}

\noindent We will now show the claimed properties of $\PMet$ by providing the following, general result which encompasses the existence of all small products (including the empty product, i.e. the terminal object) and all small coproducts (including the empty coproduct, i.e. the initial object).

\begin{proposition_apx}\label{prop:PMetcomplete}
$\PMet$ is a bicomplete category, i.e. it is complete and cocomplete.
\end{proposition_apx}
\begin{proof}
Let $D\colon I \to \PMet$ be a small diagram, $(X_i, d_i) := Di$ for each object $i \in I$. Obviously $UD\colon I \to \Set$ is also a small diagram. We show completeness and cocompleteness separately.
\medskip\\
\noindent\emph{Completeness}: Let $(f_i\colon X \to X_i)_{i \in I}$ be the limit cone to $UD$. We define the function $d:=\sup_{i\in I} d_i~ \circ (f_i \times f_i): X^2 \to \reals$ and claim that this is a pseudometric on $X$. Since all $d_i$ are pseudometrics, we immediately can derive that $d(x,x) = 0$ and $d(x,y) = d(y,x)$ for all $x,y \in X$. Moreover, for all $x,y,z\in X$:
\begin{align*}
	d(x,y) + d(y,z) &= \sup_{i\in I} d_i \big(f_i(x), f_i(y)\big) +\sup_{i\in I} d_i \big(f_i(y), f_i(z)\big)\\
		&\geq\sup_{i\in I}\Big( d_i \big(f_i(x), f_i(y)\big) + d_i \big(f_i(y), f_i(z)\big)\Big) \geq \sup_{i\in I} d_i \big(f_i(x), f_i(z)\big)  = d(x,z)\,.
\end{align*}
With this pseudometric all $f_j$ are nonexpansive functions $(X, d) \nonexpansiveTo (X_j,d_j)$. Indeed we have for all $j \in I$ and all $x,y \in X$
\begin{align*}
	d_j\big(f_j(x), f_j(y)\big) \leq \sup_{i \in I} d_i\big(f_i(x), f_i(y)\big) = d(x,y)\,.
\end{align*}
Moreover, if $\big(f_i'\colon (X',d') \nonexpansiveTo (X_i,d_i)\big)_{i \in I}$ is a cone to $D$, $(f_i'\colon X' \to X_i)_{i \in I}$ is a cone to $UD$ and hence there is a unique function $g\colon X' \to X$ in $\Set$ satisfying $f_i \circ g = f_i'$ for all $i \in I$. We finish our proof by showing that this is a nonexpansive function $(X',d') \nonexpansiveTo (X,d)$. By nonexpansiveness of the $f_i'$ we have for all $i \in I$ and all $x,y \in X'$ that $d_i(f_i'(x),f_i'(y)) \leq d'(x,y)$ and thus also 
\begin{align*}
	d\big(g(x),g(y)\big) &= \sup_{i \in I} d_i\Big(f_i\big(g(x)\big), f_i\big(g(y)\big)\Big) = \sup_{i \in I} d_i\big(f_i'(x), f_i'(y) \big)\leq \sup_{i \in I}d'(x,y) = d'(x,y)\,.
\end{align*}
We conclude that $\big(f_i\colon (X,d) \nonexpansiveTo (X_i,d_i)\big)_{i \in I}$ is a limit cone to $D$.
\medskip\\\noindent\emph{Cocompletness:} Let $(f_i\colon X_i \to X)_{i \in I}$ be the colimit co-cone from $UD$ and $M_X$ be the set of all pseudometrics $d_X\colon X^2 \to \reals$ on $X$ such that the $f_i$ are nonexpansive functions $(X_i,d_i) \nonexpansiveTo (X,d_X)$. We define $d:=\sup_{d_X \in M_X} d_X$ and claim that this is a pseudometric. Since all $d_X$ are pseudometrics, we can derive immediately that $d(x,x) = 0$ and $d(x,y) = d(y,x)$ for all $x,y \in X$. Moreover, for all $x,y,z\in X$ we have:
\begin{align*}
	d(x,y) + d(y,z) &= \sup_{d_X\in M_X} d_X(x,y) +\sup_{d_X\in M_X} d_X(y,z) \\
		&\geq\sup_{d_X\in M_X} \Big(d_X(x,y) + d_X(y,z)\Big) \geq \sup_{d_X\in M_X} d_X(x,z)   = d(x,z)\,.
\end{align*}
With this pseudometric all $f_j$ are nonexpansive functions $(X_j, d_j) \nonexpansiveTo (X,d)$. Indeed we have for all $j \in I$ and all $x,y \in X_j$
\begin{align*}
	d\big(f_j(x), f_j(y)\big) = \sup_{d_X \in M_X} d_X\big(f_j(x), f_j(y)\big) \leq \sup_{d_X \in M_X} d_j(x,y) = d_j(x,y)\,.
\end{align*}
Moreover, if $\big(f_i'\colon (X_i,d_i) \nonexpansiveTo (X',d')\big)_{i \in I}$ is a co-cone from $D$, $(f_i'\colon X_i \to X')_{i \in I}$ is a co-cone from $UD$ and hence there is a unique function $g\colon X \to X'$ in $\Set$ satisfying $g \circ f_i = f_i'$ for all $i \in I$. We finish our proof by showing that this is a nonexpansive function $(X,d) \nonexpansiveTo (X',d')$. Let $d_g:=d'\circ (g\times g)\colon X^2 \to \reals$, then it is easy to see that $d_g$ is a pseudometric on $X$. Moreover, for all $i \in I$ and all $x,y \in X_i$ we have 
\begin{align*}
	d_g\big(f_i(x), f_i(y)\big) = d'\Big(g\big(f_i(x)\big), g\big(f_i(y)\big)\Big) = d'\big(f_i'(x), f_i'(y)\big) \leq d_i(x,y)
\end{align*}
due to nonexpansiveness of $f_i'\colon (X_i,d_i) \nonexpansiveTo (X',d')$. Thus all $f_i$ are nonexpansive if seen as functions $(X_i, d_i) \nonexpansiveTo (X,d_g)$ and we have $d_g \in M_X$. Using this we observe that for all $x,y \in X$ we have
\begin{align*}
	d'\big(g(x),g(y)\big) &= d_g(x,y) \leq \sup_{d_X \in M_X} d_X(x,y) = d(x,y)
\end{align*}
which shows that $g$ is a nonexpansive function $ (X', d') \nonexpansiveTo (X,d)$. We conclude that $\big(f_i\colon (X_i,d_i) \nonexpansiveTo (X,d)\big)_{i \in I}$ is a colimit co-cone from $D$.\qedhere
\end{proof}

\begin{proposition_apx}
\label{prop:monotone}
Let $\LiftedFunctor{F}\colon \PMet \to \PMet$ be a lifting of $F\colon \Set \to \Set$. For a pseudometric space $(X,d)$ let $\LiftedMetric{F}{d}$ denote the pseudometric on $FX$ which we obtain by applying $\LiftedFunctor{F}$ to $(X,d)$. Then $\LiftedFunctor{F}$ is monotone on pseudometrics in the following sense: If we have two pseudometrics $d_1 \leq d_2$ on a common set $X$ we also have $\LiftedMetric{F}{d_1} \leq \LiftedMetric{F}{d_2}$. 
\end{proposition_apx}

\begin{proof}
Since $d_1 \leq d_2$ the identity function on the set $X$ can be regarded as a nonexpansive function $f\colon (X,d_2) \nonexpansiveTo (X, d_1)$ because we have for all $x,y \in X$ that $d_1(f(x),f(y)) = d_1(x,y) \leq d_2(x,y)$. By functoriality of $\LiftedFunctor{F}$ also $\LiftedFunctor{F}f\colon (FX, \LiftedMetric{F}{d_2}) \nonexpansiveTo (FX, \LiftedMetric{F}{d_1})$ is nonexpansive, i.e. for all $t_1, t_2 \in FX$ we have $\LiftedMetric{F}{d_1}(FUf(t_1),FUf(t_2)) \leq \LiftedMetric{F}{d_2}(t_1,t_2)$ and moreover $\LiftedMetric{F}{d_1}(FUf(t_1),FUf(t_2)) = \LiftedMetric{F}{d_1}(F\id_X(t_1),F\id_X(t_2))= \LiftedMetric{F}{d_1}(\id_{FX}(t_1),\id_{FX}(t_2) )= \LiftedMetric{F}{d_1}(t_1, t_2)$ and thus $\LiftedMetric{F}{d_1} \leq \LiftedMetric{F}{d_2}$. 
\end{proof}

\subsection{\nameref{sec:lifting}}
We show that the following definition is sound and has the claimed properties.
\restate{def:kantorovich}
\begin{proposition_apx}
\label{prop:kantorovich-is-pseudometric}
The Kantorovich pseudometric is a pseudometric on $FX$.
\end{proposition_apx}
\begin{proof}
Reflexivity and symmetry are an immediate consequence of the fact that $d_e$ is a (pseudo)metric. In order to show the triangle inequality let $t_1, t_2, t_3 \in FX$. Then we have
\begin{align*}
	& \quad \LiftedMetric{F}{d}(t_1, t_2) + \LiftedMetric{F}{d}(t_1, t_2) \\
	&=\sup_{f \colon (X,d) \nonexpansiveTo (\reals,d_e)} {d_e\left(\EvaluationFunctor{F}f(t_1),\EvaluationFunctor{F}f(t_2)\right)} + \sup_{f \colon (X,d) \nonexpansiveTo (\reals,d_e)} {d_e\left(\EvaluationFunctor{F}f(t_2),\EvaluationFunctor{F}f(t_3)\right)}\\
	&\geq \sup_{f \colon (X,d) \nonexpansiveTo (\reals,d_e)} {\left(d_e\left(\EvaluationFunctor{F}f(t_1),\EvaluationFunctor{F}f(t_2)\right) + d_e\left(\EvaluationFunctor{F}f(t_2),\EvaluationFunctor{F}f(t_3)\right)\right)}\\
	&\geq \sup_{f \colon (X,d) \nonexpansiveTo (\reals,d_e)} {d_e\left(\EvaluationFunctor{F}f(t_1),\EvaluationFunctor{F}f(t_3)\right)} = \LiftedMetric{F}{d}(t_1,t_3)
\end{align*}
where the second inequality follows from the fact that $d_e$ is a (pseudo)metric. 
\end{proof}

\begin{proposition_apx}
\label{prop:kantorovich-lifting-is-functorial}
The Kantorovich lifting $\LiftedFunctor{F}$ is a functor on pseudometric spaces.
\end{proposition_apx}
\begin{proof}
$\LiftedFunctor{F}$ preserves identities and composition of arrows because $F$ does. Moreover, it preserves nonexpansive functions: Let $f\colon (X,d_X) \nonexpansiveTo (Y,d_Y)$ be nonexpansive and $t_1, t_2 \in FX$, then
\begin{align*}
	\Kantorovich{F}{d_Y}\left(Ff(t_1), Ff(t_2)\right)&=\sup_{g \colon (Y,d_Y) \nonexpansiveTo (\reals, d_e)} d_e(\EvaluationFunctor{F}(g \circ f)(t_1),\EvaluationFunctor{F}(g \circ f)(t_2)) \\
	&\leq \sup_{h \colon (X,d_X) \nonexpansiveTo (\reals,d_e)} d_e\Big(\EvaluationFunctor{F}(h)(t_1),\EvaluationFunctor{F}(h)(t_2)\Big) = \Kantorovich{F}{d_X}(t_1, t_2)
\end{align*}
due to the fact that, of course, the composition $(g \circ f) \colon (X,d_X) \nonexpansiveTo (\reals,d_e)$ is nonexpansive.
\end{proof}

\restate{prop:kantorovich-lifting-preserves-isometries}
\begin{proof}
Let $f\colon (X,d_X) \nonexpansiveTo (Y,d_Y)$ be an isometry, i.e. a function such that $d_Y \circ (f \times f) = d_X$. Since the Kantorovich lifting $\LiftedFunctor{F}$ is a functor on pseudometric spaces, we already know that $\LiftedFunctor{F}f$ is nonexpansive, i.e. we know that $\Kantorovich{F}{d_Y} \circ ({F}f \times {F}f) \leq \LiftedMetric{F}{d_X}$ thus we only have to show the opposite inequality. We will do that by constructing for every nonexpansive function $g\colon (X, d_X) \nonexpansiveTo (\reals, d_e)$ a nonexpansive function $h\colon (Y,d_Y) \nonexpansiveTo (\reals, d_e)$ such that for every $t_1,t_2 \in FX$ we have equality $d_e(\EvaluationFunctor{F}h(Ff(t_1)), \EvaluationFunctor{F}h(Ff(t_2)))= d_e(\EvaluationFunctor{F}g(t_1), \EvaluationFunctor{F}g(t_2))$, because then we have
\begin{align*}
	\Kantorovich{F}{d_Y} \circ ({F}f \times {F}f)(t_1, t_2) &= \sup \{d_e(\EvaluationFunctor{F}h (Ff)(t_1),\EvaluationFunctor{F}h (Ff)(t_2)) \mid h\colon (Y,d_Y) \nonexpansiveTo (\reals,d_e)\}\\
	& \geq \sup \{d_e(\EvaluationFunctor{F}g (t_1),\EvaluationFunctor{F}g (t_2)) \mid g\colon (X,d_X) \nonexpansiveTo (\reals,d_e)\} \\
	&= \Kantorovich{F}{d_X}(t_1,t_2)\,.
\end{align*}
We construct $h$ as follows: For each $y \in f[X]$ we choose a fixed $x_y \in f^{-1}[\set{y}]$ and define 
\begin{align*}
	h(y) := 	\begin{cases}
					g(x_y), & y \in f[X]\\
					\inf\limits_{y' \in f[X]} h(y') + d_Y(y', y), & \text{else.}
				\end{cases}
\end{align*}
Let us first verify that this definition is independent of our choice of the $x_y$. Given $x_1, x_2 \in X$ with $f(x_1) = f(x_2) = y$ we get $d_X(x_1, x_2) = d_Y(f(x_1), f(x_2)) = d_Y(y, y) = 0$ using the fact that $f$ is an isometry. Thus by nonexpansiveness of $g$ we necessarily have $d_e(g(x_1), g(x_2)) \leq d_X(x_1, x_2) = 0$ and because $d_e$ is a metric this yields $g(x_1) = g(x_2)$. With the same reasoning we obtain $(h \circ f) (x) = h(f(x)) = g(x_{f(x)}) = g(x)$ for all $x \in X$ and therefore the desired equality
\begin{align*}
	d_e(\EvaluationFunctor{F}h(Ff(t_1)), \EvaluationFunctor{F}h(Ff(t_2))) &= d_e(\EvaluationFunctor{F}(h \circ f)(t_1), \EvaluationFunctor{F}(h \circ f)(t_2)) = d_e(\EvaluationFunctor{F}g(t_1), \EvaluationFunctor{F}g(t_2))\,.
\end{align*}
It remains to show that $h$ is nonexpansive, which we will do by distinguishing three cases. 
\begin{enumerate}
	\item Let $y_1, y_2 \in f[X]$, then there are $x_1, x_2 \in X$ with $f(x_i)=y_i$. We calculate
		\begin{align*}
			d_e(h(y_1), h(y_2)) = d_e(g(x_1), g(x_2)) \leq d_X(x_1, x_2) = d_Y(f(x_1), f(x_2)) = d_Y(y_1, y_2)
		\end{align*}
		using nonexpansiveness of $g$ and the fact that $f$ is an isometry.
	\item Let without loss of generality\footnote{For the case $y_1 \in Y \setminus f[X]$, $y_2 \in f[X]$ we can simply use symmetry of $d_e$ and $d_Y$.} $y_1 \in f[X]$ (so there is $x_1 \in X$ with $f(x_1) = y_1$) and $y_2 \in Y \setminus f[X]$, then by \lemmaref{lem:sum-vs-dist} we have the equivalence:
		\begin{align*}
			\phantom{\iff \quad} & d_e(h(y_1), h(y_2)) \leq d_Y(y_1, y_2) \\
			\iff \quad & h(y_1) \leq h(y_2) + d_Y(y_1, y_2)\quad \wedge \quad h(y_2) \leq h(y_1) + d_Y(y_1, y_2)
		\end{align*}
		We will show these inequalities separately. The second one is easy:
		\begin{align*}
			h(y_1) + d_Y(y_1, y_2) \geq \inf_{y' \in f[X]} \left(h(y') + d_Y(y', y_2)\right) = h(y_2)
		\end{align*}
		because $y_1 \in f[X]$. For the first one we calculate
		\begin{align*}
			h(y_2) + d_Y(y_1, y_2) &= \inf_{y' \in f[X]} \Big(h(y') + d_Y(y', y_2) \Big) + d_Y(y_1, y_2)\\
				&= \inf_{y' \in f[X]} \Big(h(y') + d_Y(y', y_2) + d_Y(y_2, y_1)\Big)\\
				&\geq \inf_{y' \in f[X]} \left(h(y') + d_Y(y', y_1)\right) = h(y_1)
		\end{align*}
		using symmetry and the triangle inequality for $d_Y$. Observe that the last equality is not true by definition because $y_1 \in f[X]$. Certainly we have $h(y_1) = h(y_1) +0 = h(y_1) + d_Y(y_1,y_1) \geq \inf_{y' \in f[X]} \left(h(y') + d_Y(y', y_1)\right)$.
Assume this inequality was strict, then there would be $y' \in f[X]$ such that $h(y_1) > h(y') + d_Y(y', y_1)$ which, using \lemmaref{lem:sum-vs-dist}, yields $d_e(h(y_1), h(y')) > d_Y(y_1, y')$. This contradicts nonexpansiveness of $h$ for elements of $f[X]$. Thus our assumption must have been wrong and the inequality must be an equality.
			\item Let $y_1, y_2 \in Y \setminus f[X]$. As in the previous case we use \lemmaref{lem:sum-vs-dist}, however, this time the two inequalities can be shown using exactly the same reasoning. Hence we only show the first one (which in turn is similar as in the prove above):
		\begin{align*}
			h(y_2) + d_Y(y_1, y_2) &= \inf_{y' \in f[X]} \Big(h(y') + d_Y(y', y_2) \Big) + d_Y(y_1, y_2)\\
				&= \inf_{y' \in f[X]} \Big(h(y') + d_Y(y', y_2) + d_Y(y_2, y_1)\Big)\\
				&\geq \inf_{y' \in f[X]} \Big(h(y') + d_Y(y', y_1)\Big) = h(y_1)
		\end{align*}
		The main difference to the proof above is that the last equality now holds by definition because $y_1 \not \in f[X]$. \qedhere
\end{enumerate}
\end{proof}

\restate{exa:probability-distribution-functor}
\begin{proof}
We calculate $\EvaluationFunctor{\Distributions}$. Let $g\colon X \to \reals$ be a function and $P \in \Distributions X$, then
\begin{align*}
	\EvaluationFunctor{\Distributions }g(P) &= \ev_\Distributions \circ \Distributions g (P)= \ExpectedValue{\Distributions g(P)}{\id_\reals} = \sum_{r \in \reals} r \cdot \Distributions g(P)(r) = \sum_{r \in \reals} \left(r \cdot \sum_{x \in g^{-1}[\set{r}]}P(x)\right)\\
	&=\sum_{r \in \reals}\sum_{x \in g^{-1}[\set{r}]} r \cdot P(x) = \sum_{r \in \reals}\sum_{x \in g^{-1}[\set{r}]} g(x) \cdot P(x) = \sum_{x \in X}g(x) \cdot P(x)\,.
\end{align*}
With this calculation at hand we see, for $f\colon (X,d) \nonexpansiveTo ([0,1],d_e)$ and $P_1, P_2 \in \Distributions X$ that $\EvaluationFunctor{\Distributions}f(P_1)-\EvaluationFunctor{\Distributions}f(P_2) = \sum_{x \in X} f(x) (P_1(x)-P_2(x))$.
\end{proof}

\subsection{\nameref{sec:wasserstein}}
\restate{lem:coupling}
\begin{proof}
Consider the diagrams below. The left diagram is a pullback square: Given any set $P$ along with functions $p_1,p_2\colon P \to X \times X$ satisfying the condition $\pi_2^2 \circ p_1 = \pi_1^2 \circ p_2$ the unique mediating arrow $u\colon P \to X\times X \times X$ is given by $u = \langle\pi_1^2 \circ p_1, \pi_2^2\circ p_1, \pi_2^2\circ p_2\rangle = \langle\pi_1^2 \circ p_1, \pi_1^2\circ p_2, \pi_2^2\circ p_2\rangle$. 

Now consider the right diagram. Since $F$ preserves weak pullbacks, the square in the middle of the right diagram is a weak pullback.\\
\begin{tikzpicture}
	\matrix(m)[matrix of math nodes, column sep=1pt, row sep=15pt]{
						& X \times X \times X &&&& F(X \times X \times X)\\
		X \times X 	& 								& X \times X & \hspace{1.4cm}& F(X \times X) && F(X \times X)\\
						& X && FX && FX && FX\\
	};
	\draw[->] (m-1-2) edge node[above left]{$\langle\pi_1^3, \pi_2^3\rangle$} (m-2-1);
	\draw[->] (m-1-2) edge node[above right]{$\langle\pi_2^3, \pi_3^3\rangle$} (m-2-3);
	\draw[->] (m-2-1) edge node[below left]{$\pi_2^2$} (m-3-2);
	\draw[->] (m-2-3) edge node[below right]{$\pi_1^2$} (m-3-2);
	
	\draw[->] (m-1-6) edge node[above left]{$F(\langle\pi_1^3, \pi_2^3\rangle)$} (m-2-5);
	\draw[->] (m-1-6) edge node[above right]{$F(\langle\pi_2^3, \pi_3^3\rangle)$} (m-2-7);
	\draw[->] (m-2-5) edge node[below left]{$F\pi_2^2$} (m-3-6);
	\draw[->] (m-2-7) edge node[below right]{$F\pi_1^2$} (m-3-6);
	\draw[->] (m-2-5) edge node[below right]{$F\pi_1^2$} (m-3-4);
	\draw[->] (m-2-7) edge node[below left]{$F\pi_2^2$} (m-3-8);
\end{tikzpicture}\\
By the (weak) universality of the pullback and the fact that $F\pi_2^2(t_{12}) = t_2 = F\pi_1^2(t_{23})$ we obtain\footnote{Explicitly: Consider $\set{t_2}$ with functions $p_1, p_2\colon \set{t_2}\to F(X\times X)$ where $p_1(t_2) = t_{12}$ and $p_2(t_2) = t_{23}$, then by the weak pullback property there is a (not necessarily unique) function $u\colon \set{t_2} \to F(X \times X \times X)$ satisfying $F(\langle\pi_1^3, p_2^3\rangle) \circ u = p_1$ and $F(\langle\pi_2^3, p_3^3\rangle) \circ u = p_2$. We simply define $t_{123} := u(t_2)$.} an element $t_{123} \in F(X\times X \times X)$ which satisfies the two equations of the lemma and moreover $F\pi_1^3(t_{123}) = F(\pi_1^2 \circ \langle\pi_1^3, \pi_2^3\rangle)(t_{123}) = F\pi_1^2 \circ F(\langle\pi_1^3, \pi_2^3\rangle)(t_{123}) = F\pi_1^2(t_{12}) = t_1$ and analogously $F\pi_2^3(t_{123}) = t_2$, $F\pi_3^3(t_{123}) = t_3$ yielding $t_{123} \in \Couplings{F}(t_1, t_2, t_3)$.
\end{proof}

\noindent We recall the definition of well-behavedness for an evaluation function.
\restate{def:well-behaved}
\noindent Please note that Condition~2 above can be rephrased as $d_e(\EvaluationFunctor{F}\pi_1(t),\EvaluationFunctor{F}\pi_2(t)) \le \EvaluationFunctor{F}d_e(t)$. In the following proofs we will often use this characterization.

\restate{ex:evaluation-max}
\begin{proof}
We first show that $\max\colon \PowersetFinite[0,\infty] \to [0,\infty]$ is well-behaved:
\begin{enumerate}
\item Let $f,g\colon X\to \reals$ with $f\le g$. Let $Y\in \PowersetFinite X$, i.e., $Y\subseteq X$ and $Y$ finite. Then we have $\EvaluationFunctor{\PowersetFinite}f(Y) = \ev_{\PowersetFinite}(\PowersetFinite f(Y)) = \max f[Y] \le \max g[Y] = \ev_{\PowersetFinite}(\PowersetFinite g(Y)) = \EvaluationFunctor{\PowersetFinite}g(Y)$.
\item For $T\subseteq \reals^2$ we have $\max d_e[T] \ge d_e(\max \pi_1[T],\max \pi_2[T])$. This can be shown as follows: For $T = \emptyset$ this is obviously true. Otherwise let $m_i = \max\pi_i[T]$ and let $n_1,n_2\in [0,\infty]$ such that $(m_1,n_2),(n_1,m_2)\in T$. It must hold that $m_1\ge n_1$, $m_2\ge n_2$. We distinguish the following cases: if $m_1\ge m_2$, then $m_1\ge m_2\ge n_2$, so a simple case distinction $d_e(m_1,n_2) \ge d_e(m_1,m_2)$ and hence there is a pair in $T$ with distance larger than or equal to $d_e(m_1,m_2)$. If $m_2\ge m_1$ we can conclude with an analogous argument.

\item We have $\PowersetFinite i [\PowersetFinite \{0\}] = \PowersetFinite i [\{\emptyset,\{0\}\}] = \{i[\emptyset],i[\{0\}]\} = \{\emptyset, \{0\}\} = \max^{-1}[\{0\}]$.
\end{enumerate}

\noindent The function $\min\colon \PowersetFinite(\prealinf) \to \prealinf$ is not well-behaved. It does not satisfy Condition~2, nor Condition~3: $\min d_e(T) \ge d_e(\min \pi_1(T),\min \pi_2(T))$ fails for $T = \{(0,1),(1,1)\}$ and the set $\{0,1\}$ is contained in the kernel.
\end{proof}

\restate{prop:wasserstein-is-pseudometric}
\begin{proof}
We check all three properties of a pseudometric separately.\smallskip\\
\emph{Reflexivity:} Let $t_1 \in FX$.  To show reflexivity we will construct a coupling $t \in \Couplings{F}(t_1,t_1)$ such that $\EvaluationFunctor{F}d(t) = 0$. In order to do that, let $\delta \colon X \to X^2, \delta(x) = (x,x)$ and define $t := F\delta(t_1)$. Then $F\pi_i(t) = F(\pi_i \circ \delta)(t_1) = F(\id_X)(t_1) = t_1$ and thus $t \in \Couplings{F}(t_1,t_1)$. Since $d$ is reflexive, $d \circ \delta\colon X \to [0,\top]$ is the constant zero function. Let $i\colon \set{0} \hookrightarrow [0,\top], i(0) = 0$ and for any set $X$ let $!_X\colon X \to \set{0}, !_X(x) = 0$. Then also $i \circ !_X \colon X \to [0,\top]$ is the constant zero function and thus $d \circ \delta = i \circ !_X$. Using this we conclude that $\EvaluationFunctor{F}{d}(t) = \EvaluationFunctor{F}{d}(F\delta(t_1)) = \EvaluationFunctor{F}(d \circ \delta)(t_1) =  \EvaluationFunctor{F}(i \circ !_X) = \ev_F(Fi((F!_X)(t_1))) = 0$ where the last equality follows from the fact that $F!_X(t_1) \in F\{0\}$ and $\ev_F$ is well-behaved (Condition~3 of \definitionref{def:well-behaved}).
\smallskip
\noindent\emph{Symmetry:} Let $t_1, t_2 \in FX$, $t_{12} \in \Couplings{F}(t_1,t_2)$ and $\sigma := \langle \pi_2, \pi_1\rangle$ be the swap map on $X \times X$, i.e. $\sigma\colon X \times X \rightarrow X \times X$, $\sigma(x_1,x_2) = (x_2,x_1)$. We define $t_{21} := F\sigma(t_{12}) \in F(X \times X)$ and observe that it satisfies $F\pi_1(t_{21}) = F\pi_1(F\sigma(t_{12})) = F(\pi_1 \circ \sigma)(t_{12})= F\pi_2(t_{12}) = t_2$ and analogously $F\pi_2(t_{21}) = t_1$, thus $t_{21} \in \Couplings{F}(t_2, t_1)$. Moreover, due to symmetry of $d$ (i.e. $d \circ \sigma = d$), we obtain $\EvaluationFunctor{F}d(t_{21}) = \ev_{F} \big(Fd(t_{21})\big) = \ev_F \big(Fd(F\sigma(t_{12})\big) = \ev_F\big(F(d \circ \sigma)(t_{12}) \big)= \ev_F \big(Fd(t_{12})\big) = \EvaluationFunctor{F}d(t_{12})$ which yields the desired symmetry.

\smallskip
\noindent \emph{Triangle inequality:} Using \lemmaref{lem:alt-char-triangle} we will show that for every $t_1 \in FX$ the function $\Wasserstein{F}{d}(t_1,\_)\colon (FX, \Wasserstein{F}{d}) \nonexpansiveTo (\reals,d_e)$ is nonexpansive, i.e. for all $t_2,t_3 \in FX$ we have $d_e(\Wasserstein{F}{d}(t_1,t_2),\Wasserstein{F}{d}(t_1,t_3)) \leq \Wasserstein{F}{d}(t_2,t_3)$. 

We start by observing that (also using \lemmaref{lem:alt-char-triangle}) for all $x \in X$ the function $d(x,\_)$ is nonexpansive, i.e. $d_e \circ (d \times d) \circ \langle \langle\pi_1, \pi_2\rangle, \langle\pi_1, \pi_3\rangle \rangle \leq d \circ \langle \pi_2, \pi_3\rangle$. Monotonicity of $\EvaluationFunctor{F}$ yields $\EvaluationFunctor{F}\left(d_e \circ (d \times d) \circ \langle \langle\pi_1, \pi_2\rangle, \langle\pi_1, \pi_3\rangle \rangle \right) \leq \EvaluationFunctor{F}(d \circ \langle \pi_2, \pi_3\rangle)$. Let $t_1, t_2, t_3 \in FX$ and assume there are $t_{12} \in \Couplings{F}(t_1, t_2)$ and $t_{23} \in \Couplings{F}(t_2,t_3)$. By the \nameref{lem:coupling} we get a $t_{123} \in \Couplings{F}(t_1,t_2,t_3)$ and observe that $t_{13} := F(\langle \pi_1, \pi_3\rangle)(t_{123})$ satisfies $t_{13} \in \Couplings{F}(t_1,t_3)$. Plugging in $t_{123}$ in the inequality from above yields $\EvaluationFunctor{F}d_e\left(F(d\times d)(t_{12}, t_{13})\right) \leq \EvaluationFunctor{F}d(t_{23})$. Using well-behavedness (Condition~2) of $\ev_F$ on the left hand side we obtain the following, intermediary result:
\begin{align}
	\forall t_{12} \in \Couplings{F}(t_1, t_2)\ \forall t_{23}\in \Couplings{F}(t_2,t_3)\ \exists t_{13} \in \Couplings{F}(t_1, t_3) : d_e\big(\EvaluationFunctor{F}d(t_{12}), \EvaluationFunctor{F}d(t_{13})\big) \leq \EvaluationFunctor{F}d(t_{23}) \,.\label{eq:proof-wasserstein-triangle-intermediary-result}
\end{align}
Let $d_{ij} := \Wasserstein{F}{d}(t_i,t_j)$. As explained in the beginning, we want to use \lemmaref{lem:alt-char-triangle} and thus we have to show that $d_e(d_{12}, d_{13}) \leq d_{23}$. This is obviously true for $d_{12} = d_{13}$. Without loss of generality we assume $d_{12} < d_{13}$ and claim that for all $\epsilon > 0$ we can find a coupling $t_{12} \in \Couplings{F}(t_1,t_2)$ such that for all couplings $t_{13} \in \Couplings{F}(t_1,t_3)$ we have
\begin{align}
	d_e(d_{12}, d_{13}) \leq \epsilon + d_e\big(\EvaluationFunctor{F}d(t_{12}), \EvaluationFunctor{F}d(t_{13})\big)\,.\label{eq:proof-wasserstein-triangle}
\end{align}
Since the Wasserstein distance is defined as an infimum, we have $d_{13} \leq \EvaluationFunctor{F}d(t_{13})$ and for every $\epsilon > 0$ we can pick a coupling $t_{12} \in \Couplings{F}(t_1,t_2)$, such that $\EvaluationFunctor{F}d(t_{12}) \leq d_{12} + \epsilon$. If $d_{13} = \infty$ we have $d_e(d_{12},d_{13}) = \infty$ but also $\EvaluationFunctor{F}d(t_{13}) = \infty$ and $\EvaluationFunctor{F}d(t_{12}) \leq d_{12}+ \epsilon < \infty + \epsilon = \infty$ and therefore $d_e\big(\EvaluationFunctor{F}d(t_{12}), \EvaluationFunctor{F}d(t_{13})\big) = \infty$ and thus \eqref{eq:proof-wasserstein-triangle} is valid. For $d_{13} < \infty$ we have 
\begin{align*}
d_e(d_{12}, d_{13}) &= d_{13} - d_{12} \leq \EvaluationFunctor{F}d(t_{13}) - (\EvaluationFunctor{F}d(t_{12})-\epsilon) = \epsilon + (\EvaluationFunctor{F}d(t_{13}) - (\EvaluationFunctor{F}d(t_{12}))\\ 
&\leq \epsilon +|\EvaluationFunctor{F}d(t_{13})- \EvaluationFunctor{F}d(t_{12})| \leq 
\epsilon + d_e\big(\EvaluationFunctor{F}d(t_{12}), \EvaluationFunctor{F}d(t_{13})\big)
\end{align*}
where the last inequality is due to the fact that $\EvaluationFunctor{F}d(t_{12}) < \infty$. Hence we have established our claimed validity of \eqref{eq:proof-wasserstein-triangle}. Using this, \eqref{eq:proof-wasserstein-triangle-intermediary-result} and the fact that -- as above -- given $\epsilon > 0$ we have a coupling $t_{23}$ such that $\EvaluationFunctor{F}d(t_{23}) \leq d_{23} + \epsilon$ we obtain 
\begin{align*}
	d_e(d_{12}, d_{13}) \leq \epsilon + d_e\big(\EvaluationFunctor{F}d(t_{12}), \EvaluationFunctor{F}d(t_{13})\big)\leq \epsilon + \EvaluationFunctor{F}d(t_{23}) \leq 2\epsilon + d_{23}
\end{align*}
which also proves $d_e(d_{12}, d_{13}) \leq d_{23}$. Indeed if $d_e(d_{12}, d_{13})> d_{23}$ then we would have $d_e(d_{12}, d_{13}) = d_{23}+\epsilon'$ and we just take $\epsilon < \epsilon'/2$ which yields the contradiction $	d_e(d_{12}, d_{13}) \leq 2\epsilon + d_{23} < \epsilon' + d_{23} = d_e(d_{12}, d_{13})$.

So far we have established the triangle inequality for cases where there are couplings $t_{12} \in \Couplings{F}(t_1,t_2)$ and $t_{23} \in \Couplings{F}(t_2,t_3)$. If both do not exist we have for sure $d_e(d_{12},d_{13}) \leq \infty = d_{23}$. Finally we observe that due to the \nameref{lem:coupling} it cannot be the case that just one out of three couplings does not exist as we could construct the third from the other two.
\end{proof}

\begin{proposition_apx}
\label{prop:wasserstein-lifting-is-functorial}
The Wasserstein lifting $\LiftedFunctor{F}$ is a functor on pseudometric spaces.
\end{proposition_apx}
\begin{proof}
$\LiftedFunctor{F}$ preserves identities and composition of arrows because $F$ does. Moreover, it preserves nonexpansive functions: Let $f\colon (X,d_X) \nonexpansiveTo (Y,d_Y)$ be nonexpansive and $t_1,t_2 \in FX$. Every $t \in \Couplings{F}(t_1,t_2)$ satisfies $Ff(t_i) = Ff(F\pi_i(t)) = F(f\circ \pi_i)(t) = F(\pi_i\circ (f\times f))(t) = F\pi_i(F(f\times f)(t))$. Hence we can calculate
\begin{align}
	\Wasserstein{F}{d_X}(t_1,t_2) &= \inf \{\EvaluationFunctor{F}d_X(t) \mid t \in \Couplings{F}(t_1,t_2)\}\nonumber \\
	&\geq \inf \{\EvaluationFunctor{F}d_X(t) \mid t \in F(X \times X),\ F\pi_i(F(f\times f)(t)) = Ff(t_i)\}\label{eq:wassersteinLiftingIsFunctor:proof:1}\\
	&\geq \inf \{\EvaluationFunctor{F}d_Y(F(f\times f)(t)) \mid t \in F(X \times X),\ F\pi_i(F(f\times f)(t)) = Ff(t_i)\}\label{eq:wassersteinLiftingIsFunctor:proof:2}\\
	&\geq \inf \{\EvaluationFunctor{F}d_Y(t') \mid t' \in \Gamma_F(Ff(t_1), Ff(t_2))\} =  \Wasserstein{F}{d_Y}(Ff(t_1),Ff(t_2))\,.\label{eq:wassersteinLiftingIsFunctor:proof:3}
\end{align}
In this calculation the inequality \eqref{eq:wassersteinLiftingIsFunctor:proof:1} is due to our initial observation. \eqref{eq:wassersteinLiftingIsFunctor:proof:2} holds because $f$ is nonexpansive, i.e. $d_X \geq d_Y\circ (f \times f)$ and applying the monotonicity\footnote{Condition~1 of \definitionref{def:evfct}; both $d_X$ and $d_Y \circ (f \times f)$ are functions with signature $X \times X \to \reals$} of $\EvaluationFunctor{F}$  yields $\EvaluationFunctor{F}d_X \geq \EvaluationFunctor{F}(d_Y \circ (f\times f)) = \EvaluationFunctor{F}d_Y \circ F(f\times f)$. The last inequality, \eqref{eq:wassersteinLiftingIsFunctor:proof:3}, is due to the fact that there might be more couplings $t'$ than those obtained via $F(f\times f)$.
\end{proof}

\restate{prop:wasserstein-preserves-isometries}
\begin{proof}
Let $f\colon (X,d_X)\nonexpansiveTo (Y,d_Y)$ be an isometry. Since $\LiftedFunctor{F}$ is a functor, $\LiftedFunctor{F}f$ is nonexpansive, i.e. for all $t_1,t_2\in FX$ we have $\Wasserstein{F}{d_X}(t_1,t_2) \ge \Wasserstein{F}{d_Y}(Ff(t_1),Ff(t_2))$. Now we show the opposite direction, i.e. that for all $t_1,t_2\in FX$ we have $\Wasserstein{F}{d_X}(t_1,t_2) \le \Wasserstein{F}{d_Y}(Ff(t_1),Ff(t_2))$. 

If $\Gamma_F(Ff(t_1), Ff(t_2)) = \emptyset$ we have $\Wasserstein{F}{d_Y}((Ff(t_1),Ff(t_2)) = \top \geq \Wasserstein{F}{d_X}(t_1,t_2)$. Otherwise we will construct for each coupling $t \in \Gamma_F(Ff(t_1), Ff(t_2))$ a coupling $\gamma(t) \in \Gamma_F(t_1,t_2)$ such that $\EvaluationFunctor{F}d_X(\gamma(t))=\EvaluationFunctor{F}d_Y(t)$ because then we have
\begin{align*}
	\Wasserstein{F}{d_X}(t_1,t_2) = \inf_{t'\in \Gamma_F(t_1,t_2)} \EvaluationFunctor{F}d_X(t') &\leq \inf_{t\in \Gamma_F(Ff(t_1),Ff(t_2))} \EvaluationFunctor{F}d_X(\gamma(t))\\
	&=\inf_{t\in \Gamma_F(Ff(t_1),Ff(t_2))} \EvaluationFunctor{F}d_Y(t) = \Wasserstein{F}{d_Y}(Ff(t_1), Ff(t_2))
\end{align*}
as desired. In this calculation the inequality is due to the fact that $\gamma(t) \in \Gamma_F(t_1,t_2)$. 

In order to construct $\gamma\colon \Gamma_F(Ff(t_1), Ff(t_2)) \to \Gamma_F(t_1,t_2)$, we consider the diagram below. 

\begin{center}\begin{tikzpicture}
	\matrix(m)[matrix of math nodes, column sep=40pt, row sep=15pt]{
			&				& X \times X\\
			& X \times Y & 					& Y \times X\\
		X 	&				& Y \times Y		&				& X\\
			& Y			&					& Y\\
	};
	\draw[->] (m-1-3) edge node[above left]{$\id_X \times f$} (m-2-2);
	\draw[->] (m-1-3) edge node[above right]{$f \times \id_X$} (m-2-4);
	\draw[->] (m-2-2) edge node[below left]{$f \times \id_Y$} (m-3-3);
	\draw[->] (m-2-4) edge node[below right]{$\id_Y \times f$} (m-3-3);
	\draw[->] (m-2-2) edge node[above left]{$\pi_1$} (m-3-1);
	\draw[->] (m-2-4) edge node[above right]{$\pi_2$} (m-3-5);
	\draw[->] (m-3-1) edge node[below left]{$f$} (m-4-2);
	\draw[->] (m-3-3) edge node[below right]{$\pi_1$} (m-4-2);
	
	\draw[->] (m-3-3) edge node[below left]{$\pi_2$} (m-4-4);
	\draw[->] (m-3-5) edge node[below right]{$f$} (m-4-4);
\end{tikzpicture}\end{center}

\noindent This diagram consists of pullbacks: it is easy to check that the diagram commutes. The unique mediating arrows are constructed as follows.
\begin{itemize}
	\item For the lower left part let $P$ be a set with $p_1\colon P \to X$, $p_2 \colon P \to Y \times Y$ such that $f \circ p_1 = \pi_1 \circ p_2$, then define $u\colon P \to X \times Y$ as $u :=\langle p_1, \pi_2 \circ p_2\rangle$. 
	\item Analogously, for the lower right part let $P$ be a set with $p_1\colon P \to Y \times Y$, $p_2 \colon P \to X$ such that $\pi_2 \circ p_1 = f\circ p_2$, then define $u\colon P \to X \times Y$ as $u := \langle\pi_1 \circ p_1, p_2\rangle$.
	\item Finally, for the upper part let $P$ be a set with $p_1\colon P \to X \times Y$, $p_2 \colon P \to Y \times X$ such that $(f \times \id_Y) \circ p_1 = (\id_Y \times f) \circ p_2$, then define $u\colon P \to X \times X$ as $u := \langle\pi_1 \circ p_1, \pi_2 \circ p_2\rangle$.
\end{itemize}

\noindent We apply the weak pullback preserving functor $F$ to the diagram and obtain the following diagram which hence consists of three weak pullbacks.

\begin{center}\begin{tikzpicture}
	\matrix(m)[matrix of math nodes, column sep=40pt, row sep=15pt]{
			&				& F(X \times X)\\
			& F(X \times Y) & 					& F(Y \times X)\\
		FX 	&				& F(Y \times Y)		&				& FX\\
			& FY			&					& FY\\
	};
	\draw[->] (m-1-3) edge node[above left]{$F(\id_X \times f)$} (m-2-2);
	\draw[->] (m-1-3) edge node[above right]{$F(f \times \id_X)$} (m-2-4);
	\draw[->] (m-2-2) edge node[below left]{$F(f \times \id_Y)$} (m-3-3);
	\draw[->] (m-2-4) edge node[below right]{$F(\id_Y \times f)$} (m-3-3);
	\draw[->] (m-2-2) edge node[above left]{$F\pi_1$} (m-3-1);
	\draw[->] (m-2-4) edge node[above right]{$F\pi_2$} (m-3-5);
	\draw[->] (m-3-1) edge node[below left]{$Ff$} (m-4-2);
	\draw[->] (m-3-3) edge node[below right]{$F\pi_1$} (m-4-2);
	
	\draw[->] (m-3-3) edge node[below left]{$F\pi_2$} (m-4-4);
	\draw[->] (m-3-5) edge node[below right]{$Ff$} (m-4-4);
\end{tikzpicture}\end{center}

\noindent Given a coupling $t \in \Gamma_F(Ff(t_1), Ff(t_2)) \subseteq F(Y \times Y)$ we know $F\pi_i(t) = Ff(t_i) \in FY$. 

Since the lower left square in the diagram is a weak pullback, we obtain an element\footnote{see the proof of \lemmaref{lem:coupling} for the explicit constructions} $s_1\in F(X\times Y)$ with $F\pi_1(s_1) = t_1$ and $F(f\times \id_Y)(s_1) = t$. Similarly, from the lower right square, we obtain $s_2\in F(Y\times X)$ with $F\pi_2(s_2) = t_2$ and $F(\id_Y\times f)(s_2) = t$. Again by the weak pullback property we obtain our $\gamma(t) \in F(X\times X)$ with $F(\id_X\times f)(\gamma(t)) = s_1$, $F(f\times \id_X)(\gamma(t)) = s_2$.

We convince ourselves that $\gamma(t)$ is indeed a coupling of $t_1$ and $t_2$: We have $F\pi_1(\gamma(t)) = F(\pi_1\circ (\id_X\times f))(\gamma(t)) = F \pi_1\circ F(\id_X\times f)(\gamma(t)) = F\pi_1(s_1) = t_1$ and analogously $F\pi_2(\gamma(t)) = F(\pi_2\circ (f\times\id_X))(\gamma(t)) = F\pi_2\circ F(f\times\id_X)(\gamma(t)) = F\pi_2(s_2) = t_2$.

Moreover, we have $F(f\times f)(\gamma(t)) = F((f\times \id_Y)\circ (\id_X\times f))(\gamma(t)) = F(f\times \id_Y)(s_1) = t$ and thus $\EvaluationFunctor{F}d_Y(t) = \EvaluationFunctor{F}d_Y(F(f\times f)(\gamma(t))) = \EvaluationFunctor{F}(d_Y\circ (f\times f))(t) = \EvaluationFunctor{F}(d_X(t))$ as desired. Note that the last equality is due to the fact that $f$ is an isometry.
\end{proof}

\restate{prop:preservation}
\begin{proof}
Let $(X,d)$ be a metric space and $t_1,t_2\in FX$ with $\Wasserstein{F}{d}(t_1,t_2) = 0$. We have to show that $t_1 = t_2$. Since $d$ is a metric its kernel is the set $\Delta_X = \{(x,x)\mid x\in X\}$. Hence the square on the left below is a pullback and adding the projections yields $\pi_1\circ e =\pi_2\circ e$ where $e\colon \Delta_X \hookrightarrow X\times X$ is the inclusion. Furthermore, due to Condition~3 of \definitionref{def:well-behaved}, the square on the right is a weak pullback.

\begin{center}\begin{tikzpicture}
	\matrix(m)[matrix of math nodes, column sep=30pt, row sep=15pt]{
			& \Delta_X 		& \set{0} 	& & F\set{0} & \set{0}\\
		X 	& X \times X 	& \reals	& & F\reals & \reals\\
	};
	\draw[->] (m-1-2) edge node[above]{$!_{\Delta_X}$} (m-1-3);
	\draw[right hook->] (m-1-2) edge node[left]{$e$} (m-2-2);
	\draw[right hook->] (m-1-3) edge node[right]{$i$} (m-2-3);
	\draw[->] (m-2-2) edge node[above]{$d$} (m-2-3);
	\draw[->, transform canvas={yshift=2pt}] (m-2-2) edge node[above]{$\pi_1$} (m-2-1);
	\draw[->, transform canvas={yshift=-2pt}] (m-2-2) edge node[below]{$\pi_2$} (m-2-1);
	
	% second diagram
	\draw[->] (m-1-5) edge node[above]{$!_{F\set{0}}$} (m-1-6);
	\draw[->] (m-1-5) edge node[left]{$Fi$} (m-2-5);
	\draw[right hook->] (m-1-6) edge node[right]{$i$} (m-2-6);
	\draw[->] (m-2-5) edge node[above]{$\ev_F$} (m-2-6);
\end{tikzpicture}\end{center}

\noindent Since $F$ weakly preserves pullbacks, applying it to the first diagram yields a weak pullback. By combining this diagram with the right diagram from above we obtain the diagram below where the outer rectangle is again a weak pullback. 
\begin{center}\begin{tikzpicture}
	\matrix(m)[matrix of math nodes, column sep=40pt, row sep=15pt]{
			& F\Delta_X 		& F\set{0} 	& \set{0}\\
		FX 	& F(X \times X) 	& F\reals	& \reals\\
	};
	\draw[->] (m-1-2) edge node[above]{$F!_{\Delta_X}$} (m-1-3);
	\draw[->] (m-1-2) edge node[left]{$Fe$} (m-2-2);
	\draw[->] (m-1-3) edge node[right]{$Fi$} (m-2-3);
	\draw[->] (m-2-2) edge node[above]{$Fd$} (m-2-3);
	\draw[->, transform canvas={yshift=2pt}] (m-2-2) edge node[above]{$F\pi_1$} (m-2-1);
	\draw[->, transform canvas={yshift=-2pt}] (m-2-2) edge node[below]{$F\pi_2$} (m-2-1);

	\draw[->] (m-1-3) edge node[above]{$!_{F\set{0}}$} (m-1-4);
	\draw[right hook->] (m-1-4) edge node[right]{$i$} (m-2-4);
	\draw[->] (m-2-3) edge node[above]{$\ev_F$} (m-2-4);
	
	\draw[->] (m-2-2) edge[bend right=15] node[below]{$\EvaluationFunctor{F}d$} (m-2-4);
\end{tikzpicture}\end{center}

\noindent Let $t\in F(X\times X)$ be the coupling that witnesses $\Wasserstein{F}{d}(t_1,t_2)$, i.e., $\Wasserstein{F}{d}(t_1,t_2) = \EvaluationFunctor{F}d(t) = 0$. Due to our assumption (infimum = minimum) such a coupling must exist if $\Wasserstein{F}{d}(t_1,t_2)\neq \top$. Now, since we have a weak pullback, we observe that there exists $t'\in F\Delta_X$ with $Fe(t') = t$. (Since $Fe$ is an embedding, $t'$ and $t$ actually coincide.) This implies that $t_1 = F\pi_1(t) = F\pi_1(Fe(t')) = F\pi_2(Fe(t')) = F\pi_2(t) = t_2$.
\end{proof}

\begin{proposition_apx}
\label{prop:wasserstein-upper-bound}
Let $F\colon\Set \to \Set$ be a functor with a well-behaved evaluation function $\ev_F$ and $(X,d)$ be a pseudometric space. For all $t_1, t_2 \in FX$, all couplings $t\in\Couplings{F}(t_1,t_2)$ and all nonexpansive functions $f\colon (X,d)\nonexpansiveTo (\reals,d_e)$ we have $d_e(\EvaluationFunctor{F}f(t_1),\EvaluationFunctor{F}f(t_2)) \leq \EvaluationFunctor{F}d(t)$.
\end{proposition_apx}
\begin{proof}
We have that $d\ge d_e\circ (f\times f)$ since $f$ is nonexpansive. Now, due to Condition~1 of \definitionref{def:evfct}, we obtain $\EvaluationFunctor{F}d \ge \EvaluationFunctor{F}(d_e\circ (f\times f)) = \EvaluationFunctor{F}d_e\circ F(f\times f)$. Furthermore:
\begin{align*}
	d_e(\EvaluationFunctor{F}f(t_1),\EvaluationFunctor{F}f(t_2)) 	&=	d_e(\EvaluationFunctor{F}f(F\pi_1(t)),\EvaluationFunctor{F}f(F\pi_2(t))) =  d_e(\EvaluationFunctor{F}(f\circ\pi_1)(t),\EvaluationFunctor{F}(f\circ\pi_2)(t)) \\
							&=	d_e(\EvaluationFunctor{F}(\pi_1\circ (f\times f))(t),\EvaluationFunctor{F}(\pi_2\circ (f\times f))(t)) \\
							& =  d_e(\EvaluationFunctor{F}\pi_1(F(f\times f)(t)),\EvaluationFunctor{F}\pi_2(F(f\times f)(t))) \\
							& \le \EvaluationFunctor{F}d_e(F(f\times f)(t)) = \EvaluationFunctor{F}(d_e\circ (f\times f))(t) \le \EvaluationFunctor{F}d(t)
\end{align*}
where the first inequality is due to Condition~2, the second due to the above observation.
\end{proof}
\restate{prop:wasserstein-vs-kantorovich}
\begin{proof}
This is an immediate corollary of \propositionref{prop:wasserstein-upper-bound}.
\end{proof}

\restate{exa:kantorovich-lifting-does-not-preserve-metrics2}
\begin{proof}
We prove that the evaluation function is well-behaved. Let $f,g \colon X \to [0,\infty]$ be given with $f \leq g$, then for all $t = (a, b) \in FX$ we have $\EvaluationFunctor{F}f(t) = f(a) + f(b) \leq g(a)+g(b) = \EvaluationFunctor{F}g(t)$. For $t = (r_1, r_2, r_3, r_4) \in F([0,\top]^2)$, then $\EvaluationFunctor{F}\pi_1(t) = (r_1, r_3)$, $\EvaluationFunctor{F}\pi_2(t) = (r_2, r_4)$ and thus $d_e(r_1+ r_3, r_2 + r_4) = |r_1+r_3-r_2-r_4| = |r_1-r_2 + r_3-r_4| \leq |r_1-r_2| + |r_3-r_4| = d_e(r_1, r_2) + d_e(r_3,r_4)$. Finally, $\ev_F^{-1}[\set{0}] = \{(0,0)\} = (i \times i) (\{0\} \times \{0\}) = Fi [F\{0\}]$.
\end{proof}

\restate{exa:probability-distribution-functor2}
\begin{proof} 
We just have to show well-behavedness (\definitionref{def:well-behaved}). Condition~1 is just monotonicity of the expected value. In order to prove Condition~2 we assume any probability distribution $P\colon [0,1]^2\to [0,1]$ and calculate
\begin{align*}
	&d_e\left(\EvaluationFunctor{\Distributions }\pi_1(P),\EvaluationFunctor{\Distributions }\pi_2(P)\right) = |\ExpectedValue{P}{\pi_1} - \ExpectedValue{P}{\pi_2}| = \Big|\sum_{x_1,x_2\in [0,1]} (x_1-x_2)\cdot P(x_1,x_2)\Big| \\ 	& \leq \sum_{x_1,x_2\in [0,1]} |x_1-x_2|\cdot P(x_1,x_2) = \sum_{x_1,x_2\in [0,1]} d_e(x_1,x_2)\cdot P(x_1,x_2) = \ExpectedValue{P}{d_e}=\EvaluationFunctor{\Distributions }d_e(P) \,.
\end{align*}
We denote for any set $X$ with $0 \in X$, by $\delta_0^X \in \Distributions X$ the Dirac distribution $\delta_0^X\colon X \to [0,1]$ with $\delta_0^X(0) = 1$ and $\delta_0^X(x) = 0$ for $x \in X \setminus \{0\}$. We observe that $\Distributions \set{0} = \{\delta_0^{\set{0}}\}$ and thus we can easily see that also Condition~3 holds: $\ev_\Distributions ^{-1}[\set{0}] = \{\delta_0^{[0,1]}\} = \Distributions i[\Distributions \set{0}]$. 
\end{proof}

\restate{exa:hausdorff}
\begin{proof}
We show that whenever $X_1,X_2$ are both non-empty there exists a coupling and a nonexpansive function that both witness the Hausdorff distance. Assume that the first value $\max_{x_1\in X_1} \min_{x_2\in X_2} d(x_1,x_2)$ is maximal and assume that $y_1\in X_1$ is the element of $X_1$ for which the maximum is reached. Furthermore let $y_2\in X_2$ the closest element in $X_2$, i.e.,  the element for which $d(y_1,y_2)$ is minimal. We know that for all $x_1\in X_1$ there exists $x_2^{x_1}$ such that $d(x_1,x_2^{x_1}) \le d(y_1,y_2)$ and for all $x_2\in X_2$ there exists $x_1^{x_2}$ such that $d(x_1^{x_2},x_2) \le d(y_1,y_2)$. Specifically, $x_2^{y_1} = y_2$. We use the coupling $T\subseteq X\times X$ with 
\begin{align*}
	T = \{(x_1,x_2^{x_1})\mid x_1\in X_1\} \cup \{(x_1^{x_2},x_2)\mid x_2\in X_2\}\,.
\end{align*}
Indeed we have $\PowersetFinite \pi_i (T) = X_i$ and $\PowersetFinite d(T)$ contains all distances between the elements above, of which the distance $d(y_1,y_2) = d^H(X_1,X_2)$ is maximal.
 We now define a nonexpansive function $f\colon (X,d)\to (\reals,d_e)$ as follows: $f(x) = \min_{x_2\in X_2} d(x,x_2)$.  It holds that $\max \PowersetFinite f(X_1) = \max f[X_1] = \max_{x_1\in X_1} \min_{x_2\in X_2} d(x_1,x_2) = d^H(X_1,X_2)$ and $\max \PowersetFinite f(X_2) = \max f[X_2] = 0$. Hence, the difference of both values is $d^H(X_1,X_2)$. It remains to show that $f$ is nonexpansive. Let $x,y\in X$ and let $x_2,y_2\in X_2$ be elements for which the distances $d(x,x_2),d(y,y_2)$ are minimal. Hence \begin{align*}   d(x,x_2) \le d(x,y_2) \le d(x,y) + d(y,y_2) \quad \land \quad   d(y,y_2) \le d(y,x_2) \le d(y,x) + d(x,x_2) \end{align*} Now \lemmaref{lem:sum-vs-dist} implies that $d(x,y) \ge d_e(d(x,x_2),d(y,y_2)) = d_e(f(x),f(y))$.

If $X_1=X_2=\emptyset$, we can use the coupling $T = \emptyset=\emptyset\times\emptyset$ and any function $f$. If, instead $X_1=\emptyset$, $X_2\neq\emptyset$, no coupling exists thus $\Wasserstein{F}{d} = \infty$ and we can take the constant $\infty$-function to show that also $\Kantorovich{F}{d} = \infty$ is attained.
\end{proof}

\subsection{\nameref{sec:multifunctors}}
\label{app:multifunctors}

As addition to the main text we spell out the multifunctor definitions
in details. As before, lifting is monotone on pseudometrics.
\begin{definition_apx}[Evaluation Function, Evaluation Functor, Well-Behaved]
\label{def:evfct-multi} 
An \emph{evaluation function} for a functor $F\colon \Set^n \to \Set$ is a function $\ev_F\colon F(\reals,\dots,\reals) \to \reals$. Given $\ev_F$, we define the \emph{evaluation functor} to be the functor $\EvaluationFunctor{F} \colon(\Set/\reals)^n \to \Set/\reals$ where $\EvaluationFunctor{F}(g_1,\dots,g_n) := \ev_F\circ F(g_1,\dots,g_n)$ for all $g_i \in \Set/\reals$. On arrows $\EvaluationFunctor{F}$ coincides with $F$. We call $\ev_F$ \emph{well-behaved} if it satisfies the following properties:
\begin{enumerate}
	\item $\EvaluationFunctor{F}$ is monotone: given $f_i,g_i\colon X_i\to\reals$ with $f_i\le g_i$ for all $1\leq i \leq n$, we also have $\EvaluationFunctor{F}(f_1,\dots,f_n)\le\EvaluationFunctor{F}(g_1,\dots,g_n)$.
	\item \label{cond:evfct-ge-multi} $\forall t\in F(\reals^2,\dots,\reals^2): d_e(\EvaluationFunctor{F}(\pi_1^2,\dots,\pi_1^2)(t),\EvaluationFunctor{F}(\pi_2^2,\dots,\pi_2^2)(t)) \le \EvaluationFunctor{F}(d_e,\dots,d_e)(t)$.
	\item \label{cond:evfct-zero-multi} $\ev_F^{-1}[\set{0}] = F(i,\dots,i)[F(\{0\},\dots,\{0\}]$ where $i \colon \set{0} \hookrightarrow\reals$ is the inclusion map. 
\end{enumerate}
\end{definition_apx}

\noindent The coupling definition for multifunctors is technically a bit more complicated than in the endofunctor setting but captures exactly the same idea as before.

\begin{definition_apx}[Coupling]
Let $F \colon \Set^n \to \Set$ be a functor and $m \in \N$. Given sets $X_1,\dots,X_n$ and elements $t_j \in F(X_1,\dots,X_n)$ for $1 \leq j \leq m$ we call an element $t \in F(X_1^m,\dots, X_n^m)$ such that $F(\pi^m_{1,j},\dots,\pi^m_{n,j})(t) = t_j$ a \emph{coupling} of the $t_j$ (with respect to $F$) where $\pi_{i,j}^m$ are the projections $\pi_{i,j}^m \colon X_i^m \to X_i$. We write $\Couplings{F}(t_1, t_2, \dots, t_m)$ for the set of all these couplings.\end{definition_apx}

\noindent Note that the multifunctor approach is almost identical to the endofunctor approach and the only difference is that we start with $n$ pseudometric spaces $(X_1,d_1), \dots, (X_n, d_n)$ instead of just one. Due to this we can straightforwardly adopt the proofs of the endofunctor cases to this setting. We will provide one exemplary calculation here. As in the endofunctor setting we have a gluing lemma.
\begin{lemma_apx}[Gluing Lemma for Multifunctors]
\label{lem:coupling-multi}
Let $F \colon \Set^n \to \Set$ be a weak pullback preserving multifunctor, $X_1,\dots, X_n$ sets, $t_1, t_2, t_3 \in F(X_1,\dots, X_n)$, $t_{12} \in \Couplings{F}(t_1,t_2)$, and $t_{23} \in \Couplings{F}(t_2,t_3)$ be couplings. Then there is a coupling $t_{123} \in \Couplings{F}(t_1, t_2, t_3)$ such that $F(\langle \pi_1^3, \pi_2^3\rangle)(t_{123}) = t_{12}$ and $F(\langle\pi_2^3, \pi_3^3\rangle)(t_{123}) = t_{23}$.
\end{lemma_apx}
\begin{proof}
Exactly as in the proof of the (endofunctor)~\nameref{lem:coupling}, \lemmaref{lem:coupling}, we can see that the following is a pullback square for all $1 \leq i \leq n$.
\begin{center}\begin{tikzpicture}
	\matrix(m)[matrix of math nodes, column sep=20pt, row sep=8pt]{
						& X_i^3 \\
		 X_i^2, 	& 								&  X_i^2\\
						& X_i\\
	};
	\draw[->] (m-1-2) edge node[above left]{$\langle\pi_{i,1}^3, \pi_{i,2}^3\rangle$} (m-2-1);
	\draw[->] (m-1-2) edge node[above right]{$\langle\pi_{i,2}^3, \pi_{i,3}^3\rangle$} (m-2-3);
	\draw[->] (m-2-1) edge node[below left]{$\pi_{i,2}^2$} (m-3-2);
	\draw[->] (m-2-3) edge node[below right]{$\pi_{i,1}^2$} (m-3-2);
\end{tikzpicture}\end{center}

\noindent Now consider the following diagram where we write $(\dots, X_i,\dots)$ instead of $(X_1,\dots, X_n)$ for readability. Since $F$ preserves weak pullbacks, the square in the middle of this diagram is a weak pullback.
\begin{center}\begin{tikzpicture}
	\matrix(m)[matrix of math nodes, column sep=2pt, row sep=20pt]{
			&& F(\dots, X_i^3, \dots) \\
		  &F(\dots, X_i^2, \dots) 	& 								&   F(\dots, X_i^2, \dots)\\
			F(\dots, X_i, \dots) &			&  F(\dots, X_i, \dots) && F(\dots, X_i, \dots)\\
	};
	\draw[->] (m-1-3) edge node[above left]{$ F(\dots, \langle\pi_{i,1}^3, \pi_{i,2}^3\rangle, \dots)$} (m-2-2);
	\draw[->] (m-1-3) edge node[above right]{$ F(\dots, \langle\pi_{i,2}^3, \pi_{i,3}^3\rangle, \dots)$} (m-2-4);
	\draw[->] (m-2-2) edge node[below left]{$ F(\dots, \pi_{i,2}^2, \dots)$} (m-3-3);
	\draw[->] (m-2-4) edge node[below right]{$ F(\dots, \pi_{i,1}^2, \dots)$} (m-3-3);
	
	\draw[->] (m-2-2) edge[bend right=10] node[above left]{$ F(\dots, \pi_{i,1}^2, \dots)$} (m-3-1);
	\draw[->] (m-2-4) edge[bend left=10] node[above right]{$ F(\dots, \pi_{i,2}^2, \dots)$} (m-3-5);
\end{tikzpicture}\end{center}By the (weak) universality of the pullback and the fact that 
\begin{equation*}
	F(\dots,\pi_{i,2}^2,\dots)(t_{12}) = t_2 = F(\dots,\pi_{i,1}^2,\dots)(t_{23})
\end{equation*}
we obtain an element $t_{123} \in F(\dots,X_i^3,\dots)$ which satisfies the two equations of the lemma and moreover 
\begin{align*}
	F(\dots,\pi_{i,1}^3,\dots)(t_{123}) &= F(\dots, \pi_{i,1}^2 \circ \langle\pi_{i,1}^3, \pi_{i,2}^3\rangle,\dots)(t_{123}) \\
	&= F(\dots,\pi_{i,1}^2,\dots) \circ F(\dots, \langle\pi_{i,1}^3, \pi_{i,2}^3\rangle,\dots)(t_{123}) \\
	&= F(\dots,\pi_{i,1}^2,\dots)(t_{12}) = t_1
\end{align*}
and analogously $F(\dots,\pi_{i,2}^3,\dots)(t_{123}) = t_2$, $F(\dots,\pi_{i,3}^3,\dots)(t_{123}) = t_3$ so that indeed $t_{123} \in \Couplings{F}(t_1, t_2, t_3)$ is a coupling.
\end{proof}

\noindent With these definitions at hand it is now easy to generalize our approach to multifunctors.

\begin{definition_apx}[Kantorovich/Wasserstein Pseudometric/Lifting]
\label{def:kantorovich-wasserstein-multi}
Let $F\colon \Set^n\to \Set$ be a functor with evaluation function $\ev_F$ and $(X_1,d_1), \dots, (X_n, d_n)$ be pseudometric spaces.
\begin{enumerate}
	\item The \emph{Kantorovich pseudometric} is $\KantorovichMulti{F}{d}{1,\dots,n}\colon \big(F(X_1,\dots,X_n)\big)^2\to \reals$, where
	\begin{align*}
		\KantorovichMulti{F}{d}{1,\dots,n}(t_1,t_2) := \sup_{f_i\colon (X_i,d_i) \nonexpansiveTo (\reals,d_e)} d_e\Big(\EvaluationFunctor{F}(f_1,\dots,f_n)(t_1),\EvaluationFunctor{F}(f_1,\dots,f_n)(t_2)\Big)\,.
		\end{align*}
	\item Whenever $F$ preserves weak pullbacks and $\ev_F$ is a well-behaved evaluation function, the \emph{Wasserstein pseudometric} is $\WassersteinMulti{F}{d}{1,\dots,n}\colon \big(F(X_1,\dots,X_n)\big)^2\to \reals$, where 
	\begin{align*}
		\WassersteinMulti{F}{d}{1,\dots,n}(t_1, t_2) := \inf_{t \in \Couplings{F}(t_1,t_2)} \EvaluationFunctor{F}(d_1,\dots,d_n)(t).\,
		\end{align*} \label{def:wasserstein-multi}
\end{enumerate}
The \emph{Kantorovich lifting} [\,\emph{Wasserstein lifting}\,] of $F$ is the functor $\LiftedFunctor{F}\colon \PMet^n\to\PMet$, $\LiftedFunctor{F}((X_1,d_1),\dots, (X_n,d_n)) = (F(X_1,\dots,X_n),d)$ with $d = \KantorovichMulti{F}{d}{1,\dots,n}$ [\,$d=\WassersteinMulti{F}{d}{1,\dots,n}$\,], $\LiftedFunctor{F}f = Ff$. 
\end{definition_apx}

\begin{proposition_apx}
The Kantorovich lifting for $F\colon \Set^n \to \Set$ is well defined, i.e.:
\begin{enumerate}
	\item $\KantorovichMulti{F}{d}{1,\dots,n}$ is a pseudometric on $F(X_1,\dots,X_n)$.
	\item $\LiftedFunctor{F}$ is an endofunctor on $\PMet$.
\end{enumerate}
\end{proposition_apx}
\begin{proof}
The proofs of \propositionref{prop:kantorovich-is-pseudometric} and \propositionref{prop:kantorovich-lifting-is-functorial} can easily be adapted.
\end{proof}

\begin{proposition_apx}
The Wasserstein lifting for $F\colon \Set^n \to \Set$ is well defined, i.e.:
\begin{enumerate}
	\item $\WassersteinMulti{F}{d}{1,\dots,n}$ is a pseudometric on $F(X_1,\dots,X_n)$.
	\item $\LiftedFunctor{F}$ is an endofunctor on $\PMet$.
\end{enumerate}
\end{proposition_apx}
\begin{proof}
The proofs of \propositionref{prop:wasserstein-is-pseudometric} and \propositionref{prop:wasserstein-lifting-is-functorial} can easily be adapted.
\end{proof}

\begin{proposition_apx}
\label{prop:multifunctor-lifting-properties}
Let $F\colon \Set^n \to \Set$ be a functor with evaluation function $\ev_F$.
\begin{enumerate}
	\item Let $d_i$ be pseudometrics on $X_i$, then $\KantorovichMulti{F}{d}{1,\dots,n} \leq \WassersteinMulti{F}{d}{1,\dots,n}$.
	\item Both liftings preserve isometries.
	\item Let all $d_i$ be metrics. If the infimum in \definitionref{def:kantorovich-wasserstein-multi}.\ref{def:wasserstein-multi} is a minimum for all $t_1,t_2 \in F(X_1, \dots, X_n)$ with $\KantorovichMulti{F}{d}{1,\dots,n}(t_1,t_2) = 0$ then $\KantorovichMulti{F}{d}{1,\dots,n}$ is a metric.
\end{enumerate}
\end{proposition_apx}
\begin{proof}~
\begin{enumerate}
	\item We can state and prove a multifunctor equivalent of \propositionref{prop:wasserstein-upper-bound} and as in that case our result is an immediate corollary.
	\item Again, the same arguments as in Propositions~\ref{prop:kantorovich-lifting-preserves-isometries} and~\ref{prop:wasserstein-preserves-isometries}.
\item See \propositionref{prop:preservation}. \qedhere
\end{enumerate}
\end{proof}

\noindent Whenever the two pseudometrics coincide for a functor and an evaluation function, we say that the \emph{Kantorovich-Rubinstein duality} (for multifunctors) holds.

\restate{exa:product-bifunctor}
\begin{proof}
$F$ preserves weak pullbacks: If we have two weak pullbacks in $\Set$ as indicated in the left of the diagram below, then obviously also the right diagram is a weak pullback. 
\begin{center}\begin{tikzpicture}
	\matrix(m)[matrix of math nodes, column sep=50pt, row sep=15pt]{
		P_i & X_i && P_1 \times P_2 & X_1 \times X_2\\
		Y_i & Z_i && Y_1 \times Y_2 & Z_1 \times Z_2\\
	};
	\draw[->] (m-1-1) edge node[above]{$p_i^X$} (m-1-2);
	\draw[->] (m-1-1) edge node[left]{$p_i^Y$} (m-2-1);
	\draw[->] (m-1-2) edge node[right]{$f_i$} (m-2-2);
	\draw[->] (m-2-1) edge node[below]{$g_i$} (m-2-2);
	
	\draw[->] (m-1-4) edge node[above]{$p_1^X \times p_2^X$} (m-1-5);
	\draw[->] (m-1-4) edge node[left]{$p_1^Y \times p_2^Y$} (m-2-4);
	\draw[->] (m-1-5) edge node[right]{$f_1 \times f_2$} (m-2-5);
	\draw[->] (m-2-4) edge node[below]{$g_1 \times g_2$} (m-2-5);
\end{tikzpicture}\end{center}

\noindent We proceed by checking all three conditions for well-behavedness:
\begin{enumerate}
	\item Let $f_1,f_2,g_1,g_2\colon X_i\to\reals$ with $f_1\le g_1$, $f_2\le g_2$ then for any $\mathbf{x} = (x_1,x_2)\in F(X_1,X_2) = X_1\times X_2$ we see that in the case of the maximum we have
\begin{align*}
	\EvaluationFunctor{F}(f_1, f_2)(\mathbf{x}) = \max (f_1(x_1),f_2(x_2)) \le \max (g_1(x_1),g_2(x_2)) = \EvaluationFunctor{F}(g_1, g_2)(\mathbf{x})
\end{align*}
and for the second evaluation function, we also obtain 
\begin{align*}
	\EvaluationFunctor{F}(f_1, f_2)(\mathbf{x}) &= \Big(c_1\cdot f_1^p(x_1) + c_2\cdot f_2^p(x_2)\Big)^{1/p} \\
	&\le \Big(c_1\cdot g_1^p(x_1) + c_2\cdot g_2^p(x_2)\Big)^{1/p} = \EvaluationFunctor{F}(g_1, g_2)(\mathbf{x})
\end{align*}
due to monotonicity of all involved functions since $c_1,c_2>0$.

\item Let $t:=\big((x_1,x_2),(y_1,y_2)\big)\in F(\reals^2,\reals^2) = \reals^2\times \reals^2$. We have to show $d_e\big(\EvaluationFunctor{F}(\pi_1,\pi_1)(t), \EvaluationFunctor{F}(\pi_2, \pi_2)(t)) \leq \EvaluationFunctor{F}(d_e,d_e)(t)$.
If we define $z_i = \ev_F(x_i,y_i)$ (where $\ev_F= \max$ or $\ev_F=\rho$ respectively) for $i \in \set{0,1}$ then $d_e\big(\EvaluationFunctor{F}(\pi_1,\pi_1)(t), \EvaluationFunctor{F}(\pi_2, \pi_2)(t)) = d_e(z_1,z_2)$ and $\EvaluationFunctor{F}(d_e,d_e)(t) = \ev_F(d_e(x_1,x_2), d_e(y_1,y_2))$. We thus have to show the inequality 
\begin{equation}
	d_e(z_1,z_2) \leq \ev_F(d_e(x_1,x_2), d_e(y_1,y_2))\,.\label{eq:condtwo:product}
\end{equation}
If $z_1=z_2$ this is obviously true because $d_e(z_1, z_2) = 0$ and the rhs is non-negative. We now assume $z_1 > z_2$ (the other case is symmetrical). For $\infty = z_1 > z_2$ the inequality holds because then $x_1 = \infty$ or $y_1 = \infty$ and $x_2,y_2 < \infty$ (otherwise we would have $z_2 = \infty$) so both lhs and rhs are $\infty$. Thus we can now restrict to $\infty > z_1 > z_2$ where necessarily also $x_1, y_1, x_2, y_2 < \infty$ (otherwise we would have $z_1 = \infty$ or $z_2=\infty$). According to \lemmaref{lem:sum-vs-dist}, the inequality~\eqref{eq:condtwo:product} is equivalent to showing the two inequalities
\begin{align*}
	z_1 \leq z_2 + \ev_F(d_e(x_1,x_2), d_e(y_1,y_2))\quad \land \quad z_2 \leq z_1 + \ev_F(d_e(x_1,x_2), d_e(y_1,y_2))\,.
\end{align*}
By our assumption ($\infty > z_1>z_2$) the second of these inequalities is satisfied, so we just have to show the first. We do this separately for the different evaluation functions.
\begin{itemize}
	\item Suppose $\ev_F = \max$, then $z_i = \max(x_i, y_i)$. If $z_1 = x_1$ we have 
	\begin{equation*}
		z_2 + \max\big(d_e(x_1, x_2), d_e(y_1,y_2)\big) \geq z_2 + d_e(z_1,x_2) = z_2 + (z_1-x_2) = z_1 + (z_2-x_2) \geq z_1
		\end{equation*}
	because $z_2 = \max(x_2,y_2) \geq x_2$ and therefore $(z_2-x_2) \geq 0$. The same line of argument can be applied if $z_1 = y_1$.

\item Suppose $\ev_F=\rho$. We define
\begin{equation*}
	a_1 := c_1^{1/p}x_2, \quad a_2 := c_2^{1/p}y_2, \quad b_1 := c_1^{1/p} \cdot (x_1 - x_2), \quad b_2 := c_2^{1/p} \cdot (y_1 - y_2)\,.
\end{equation*}
The Minkowski inequality tells us that
\begin{equation*}
	\big(|a_1|^p+|a_2|^p\big)^{1/p} + \big(|b_1|^p+|b_2|^p\big)^{1/p} \geq \big(|a_1 + b_1|^p + |a_2+b_2|^p\big)^{1/p}\,.
\end{equation*}
We calculate the different parts:
\begin{align*}
	(|a_1|^p + |a_2|^p)^{1/p} &= (c_1x_2^p+c_2y_2^p)^{1/p} = z_2\\
	(|b_1|^p + |b_2|^p)^{1/p} &= (c_1|x_1-x_2|^p+c_2|y_1-y_2|^p)^{1/p} = \rho(d_e(x_1,x_2), d_e(y_1,y_2))\\
	|a_1+b_1|^p &= |c_1^{1/p}x_2 + c_1^{1/p} \cdot (x_1 - x_2)|^p = c_1|x_1|^p = c_1x_1^p\\
	|a_2+b_2|^p &= |c_2^{1/p}y_2 + c_2^{1/p} \cdot (y_1 - y_2)|^p = c_2|y_1|^p = c_2y_1^p
\end{align*}
and thus the Minkowski inequality yields
\begin{equation*}
	z_2 + \rho(d_e(x_1,x_2), d_e(y_1,y_2)) \geq (c_1x_1^p + c_2y_1^p)^{1/p} = z_1
\end{equation*}
which concludes the proof.
\end{itemize}

\item Both evaluation functions satisfy Condition~\ref{cond:evfct-zero-multi} of \definitionref{def:evfct-multi}: $F(i,i)[F(\set{0},\set{0})] = (i\times i)[\set{0} \times \set{0}] = \set{(0,0)}$ and $\max^{-1}[\set{0}] = \set{(0,0)} = \rho^{-1}[\set{0}]$.
\end{enumerate}

\noindent We now prove (for both evaluation functions) that the product functor satisfies the Kantorovich-Rubinstein duality and simultaneously that the supremum (in the Kantorovich pseudometric) is a maximum and the infimum (of the Wasserstein pseudometric) is a minimum. Let $(X_1,d_1)$, $(X_1,d_2)$ be pseudometric spaces and let $t_1 = (x_1,x_2), t_2 = (y_1,y_2) \in F(X_1,X_2) = X_1 \times X_2$ be given. We define $t := ((x_1,y_1),(x_2,y_2)) \in F(X_1^2,X_2^2)$ and observe that $F(\pi_1,\pi_1)(t) = t_1$, $F(\pi_2,\pi_2)(t) = t_2$ and thus $t \in \Gamma_F(t_1,t_2)$ is a coupling. In the following we will construct nonexpansive functions $f_i\colon(X_i,d_i) \nonexpansiveTo (\reals,d_e)$ such that $d_e(\EvaluationFunctor{F}(f_1,f_2)(t_1), \EvaluationFunctor{F}(f_1,f_2)(t_2) = \EvaluationFunctor{F}(d_e,d_e)(t)$. Due to \propositionref{prop:multifunctor-lifting-properties} we can then conclude that duality holds and both supremum and infimum are attained.
\begin{itemize}
\item Suppose $\ev_F = \max$ then we have $\EvaluationFunctor{F}(d_1,d_2)(t) = \max(d_1(x_1,y_1),d_2(x_2,y_2))$ and assume wlog that $d_1(x_1,y_1)$ is the maximal element and define $f_1:=d_1(x_1,\_)$, which is nonexpansive due to \lemmaref{lem:alt-char-triangle}, and $f_2$ is the constant zero-function which is obviously nonexpansive. Then we have:
\begin{align*}
	& d_e\big(\EvaluationFunctor{F}(f_1,f_2)(t_1),\EvaluationFunctor{F}(f_1,f_2)(t_2)\big) = d_e\big(\max(f_1(x_1),f_2(x_2))),\max(f_1(y_1),f_2(y_2)\big)\\
	& = d_e\big(\max(f_1(x_1),0),\max(f_1(y_1),0)\big) = d_e\big(f_1(x_1),f_1(y_1)\big) = d_1(x_1,y_1) \\
	& = \max\big(d_1(x_1,y_1),d_2(x_2,y_2)\big) = \EvaluationFunctor{F}(d_1,d_2)(t)
\end{align*}
The case where $d_2(x_2,y_2)$ is the maximal element is treated analogously.
\item If $\ev_F = \rho$ we define $f_1 := d_1(x_1,\_)$ and $f_2:=d_2(x_2,\_)$ (which are again nonexpansive by \lemmaref{lem:alt-char-triangle}) and obtain
\begin{align*}
	& d_e\big(\EvaluationFunctor{F}(f_1,f_2)(t_1),\EvaluationFunctor{F}(f_1,f_2)(t_2)\big)\\
	& =d_e\left(\left(c_1f_1^p(x_1)+c_2f_2^p(x_2)\right)^{1/p},\left(c_1f_1^p(y_1)+c_2f_2^p(y_2)\right)^{1/p}\right)\\
	& = d_e\left(0, \left(c_1d_1^p(x_1, y_1)+c_2d_2^p(x_2,y_2)\right)^{1/p}\right) \\
	&= \left(c_1d_1^p(x_1, y_1)+c_2d_2^p(x_2,y_2)\right)^{1/p} = \EvaluationFunctor{F}(d_1,d_2)(t)
\end{align*}
which completes the proof.\qedhere
\end{itemize}
\end{proof}

\restate{exa:coproduct-bifunctor}
\begin{proof}
$F$ preserves weak pullbacks: If we have two weak pullbacks in $\Set$ as indicated in the left of the diagram below, then obviously also the right diagram is a weak pullback. 
\begin{center}\begin{tikzpicture}
	\matrix(m)[matrix of math nodes, column sep=50pt, row sep=15pt]{
		P_i & X_i && P_1 + P_2 & X_1 + X_2\\
		Y_i & Z_i && Y_1 + Y_2 & Z_1 + Z_2\\
	};
	\draw[->] (m-1-1) edge node[above]{$p_i^X$} (m-1-2);
	\draw[->] (m-1-1) edge node[left]{$p_i^Y$} (m-2-1);
	\draw[->] (m-1-2) edge node[right]{$f_i$} (m-2-2);
	\draw[->] (m-2-1) edge node[below]{$g_i$} (m-2-2);
	
	\draw[->] (m-1-4) edge node[above]{$p_1^X + p_2^X$} (m-1-5);
	\draw[->] (m-1-4) edge node[left]{$p_1^Y + p_2^Y$} (m-2-4);
	\draw[->] (m-1-5) edge node[right]{$f_1 + f_2$} (m-2-5);
	\draw[->] (m-2-4) edge node[below]{$g_1 + g_2$} (m-2-5);
\end{tikzpicture}\end{center}

\noindent We now show that the evaluation function is well-behaved. 
\begin{enumerate}
\item Let $f_1,f_2,g_1,g_2\colon X\to\reals$ with $f_1\le g_1$, $f_2\le g_2$ and $(z,i)\in F(X_1,X_2) = X_1+ X_2$. We have $\EvaluationFunctor{F}f(z,i) = \ev_F(F(f_1,f_2)(z,i)) = f_i(z) \le g_i(z) = \ev_F(F(g_1,g_2)(z,i)) = \EvaluationFunctor{F}g(z,i).$ 
\item Let $t = ((x,y),i)\in F(\reals^2,\reals^2) = \reals^2\times\{1,2\}$. Then we even obtain equality: 
\begin{align*}
	\EvaluationFunctor{F}(d_e,d_e)(t) &= \ev_F(d_e(x,y),i) = d_e(x,y) \\
	&= d_e(\ev_F(x,i),\ev_F(y,i)) = d_e(\EvaluationFunctor{F}(\pi_1,\pi_1)(t),\EvaluationFunctor{F}(\pi_2,\pi_2)(t))\,.
\end{align*}
\item Let $i\colon {0} \hookrightarrow \reals$ be the inclusion function. We have $Fi[F(\set{0},\set{0})] = (i + i) [\set{0} + \set{0}] =\set{0} \times \set{1,2} = \ev_F^{-1}[\set{0}]$.
\end{enumerate}
Now we show that the pair of functor and evaluation function $\ev_F$ satisfies the Kantorovich-Rubinstein duality and simultaneously that the supremum (in the Kantorovich pseudometric) is a maximum and the infimum (of the Wasserstein pseudometric) is a minimum iff there exists a coupling of the two given elements. Let $(X_1,d_1)$, $(X_1,d_2)$ be pseudometric spaces, $t_1 = (z,i), t_2 = (z',i') \in F(X_1,X_2) = X_1 + X_2$. Suppose $i=i'$, when we define $t = ((z,z'),i)$ and observe that $F(\pi_1,\pi_1)((z,z'),i) = t_1$, $F(\pi_2,\pi_2)((z,z'),i) = t_2$, thus $t \in \Gamma_F(t_1,t_2)$. Furthermore $\EvaluationFunctor{F}(d_1,d_2)(t) = d_i(z,z')$. If $i=i'=1$ we define $f_1:=d_1(z,\_)\colon (X_1,d_1) \nonexpansiveTo (\reals,d_e)$ which is nonexpansive according to \lemmaref{lem:alt-char-triangle} and consider an arbitrary nonexpansive function $f_2\colon (X_2,d_2) \nonexpansiveTo (\reals,d_e)$ (e.g. the constant zero-function). Then we have: 
\begin{align*}
	d_e(\EvaluationFunctor{F}(f_1,f_2)(t_1),\EvaluationFunctor{F}(f_1,f_2)(t_2)) &= d_e(\EvaluationFunctor{F}(f_1,f_2)(z,1),\EvaluationFunctor{F}(f_1,f_2)(z',1)) =  d_e(f_1(z),f_1(z')) \\
	&= d_e(0, d_1(z,z')) = d_1(z,z') = d_i(z,z')
\end{align*} 
The case $i=i'=2$ is analogous. In the case where $i\neq i'$, there is no coupling that projects to $(z,i)$ and $(z',i')$, thus $\WassersteinMulti{F}{d}{1,2}(t_1,t_2) = \top$. We show that also $\KantorovichMulti{F}{d}{1,2}(t_1,t_2) = \top$. We define $f_1$ to be the constant zero-function and $f_2$ the constant $\top$-function. We have: \begin{align*} 
	d_e(\EvaluationFunctor{F}(f_1,f_2)(t_1),\EvaluationFunctor{F}(f_1,f_2)(t_2)) &= d_e(\EvaluationFunctor{F}(f_1,f_2)(z,i),\EvaluationFunctor{F}(f_1,f_2)(z',j))\\
		& =  d_e(f_i(z),f_j(z')) = d_e(0,\top) = \top
\end{align*}
which completes the proof.
\end{proof}

\subsection{\nameref{sec:final-coalgebra}}
\restate{thm:final-coalgebra}
\begin{proof}
It can be easily shown that each of the $d_i$ is a pseudometric, since
the supremum of pseudometrics is again a pseudometric. Since
$d_\theta$ is a fixed-point, $\kappa$ is an isometry and hence
nonexpansive. Furthermore the chain converges once we reach an ordinal
whose cardinality is larger than the cardinality of the lattice of
metrics on $\Omega$.

Let $\alpha\colon (X,d)\nonexpansiveTo \LiftedFunctor{F}(X,d)$ be any $\LiftedFunctor{F}$-coalgebra, with underlying $F$-coalgebra $\alpha\colon X\to FX$ in $\Set$. Since $\kappa$ is the final $F$-coalgebra, there exists a unique function $f\colon X\to\Omega$ such that $\kappa\circ f = Ff\circ \alpha$. It is left to show that $f$ is nonexpansive function $(X,d)\nonexpansiveTo (\Omega,d_\theta)$. 

For each ordinal $i$ we define a pseudometric $e_i\colon X\times X\to \reals$ as follows: $e_0$ is the constant zero-pseudometric, $e_{i+1} := \LiftedMetric{F}{e_i}\circ (\alpha\times \alpha)$ and $e_j := \sup_{i<j} e_i$ if $j$ is a limit ordinal. We show that $e_i \le d$: Obviously $e_0 \le d$ and furthermore $e_{i+1} = \LiftedMetric{F}{e_i}\circ (\alpha\times \alpha) \le \LiftedMetric{F}{d}\circ (\alpha\times \alpha) \le d$ where the first inequality is due to the fact that the lifting preserves the order on pseudometrics and the second is nonexpansiveness of $\alpha$. If we take the limit $e_j = \sup_{i<j} e_i$, we know that $e_i\le d$ for each $i<j$ and hence also $e_j\le d$.

As an auxiliary step we will prove that all $f\colon (X,e_i)\to (\Omega,d_i)$ are nonexpansive. This holds for $i=0$ since for all $x,y\in X$ we have $e_0(x,y) = 0 = d_0(f(x),f(y))$. For $i+1$ we have
\begin{align*}
	d_{i+1}(f(x),f(y)) & = \LiftedMetric{F}{d_i}((\kappa \circ f)(x),(\kappa \circ f)(y)) = \LiftedMetric{F}{d_i}((Ff \circ \alpha)(x),(Ff \circ \alpha)(y)) \\
		& \le \LiftedMetric{F}{e_i}(\alpha(x),\alpha(y)) = e_{i+1}(x,y)\,.
\end{align*}
The inequality above holds since if $f\colon (X,e_i)\nonexpansiveTo (\Omega,d_i)$ is a nonexpansive function also $\LiftedFunctor{F}f\colon (FX,\LiftedMetric{F}{e_i})\nonexpansiveTo (F\Omega,\LiftedMetric{F}{d_i})$ is nonexpansive. Whenever $j$ is a limit ordinal we obtain:
\begin{align*}
	d_j(f(x),f(y)) & = \sup_{i<j} d_i(f(x),f(y)) \le \sup_{i<j} e_i(x,y) = e_j(x,y)\,.
\end{align*}
Finally, we combine this result with the result from above to obtain the inequality $d_\theta(f(x),f(y))  \le e_\theta(x,y) \le d(x,y)$ which shows that $f\colon (X,d)\nonexpansiveTo (\Omega,d)$ is nonexpansive.
\end{proof}

\noindent We next prove that the Wasserstein/Kantorovich liftings induced by the
finite powerset functor and by the probability distribution functor
with finite support are $\omega$-continuous.

\begin{proposition_apx}[$\omega$-continuity of the liftings of $\PowersetFinite$ and $\Distributions $]
\label{prop:w-continuity}
Let $F$ be the finite powerset functor $\PowersetFinite$ or the distribution functor $\Distributions $ (with finite support). For any set $X$ the function $\_^F$ mapping a metric $d$ over $X$ to the metric
  $\LiftedMetric{F}{d}$ over $FX$ is $\omega$-continuous, namely for any increasing
  chain of metrics $(d_i)_{i \in \N}$ over $X$, we have
  $\LiftedMetric{F}{(\sup_i d_i)} = \sup_i \LiftedMetric{F}{d_i}$.
\end{proposition_apx}

\begin{proof}
  Let $X$ be a fixed set. By \propositionref{prop:monotone}, we know
  that for any lifting, the function $\_^F$ is monotone, namely,
  whenever $d_1 \leq d_2$ it holds that $\LiftedMetric{F}{d_1} \leq \LiftedMetric{F}{d_2}$.

  Given an increasing chain of metrics $(d_i)_{i \in \N}$ over
  $X$, simply by monotonicity of $\_^F$ we deduce that
  \begin{center}
    $\sup_i d_i^F \leq (\sup_i d_i)^F$
  \end{center} 
  In fact, for any $i \in \N$, it holds that $d_i \leq \sup_i
  d_i$, hence $d_i^F \leq (\sup_i
  d_i)^F$ and therefore we conclude.

  We next prove that for the Wasserstein/Kantorovich liftings of
  either $\PowersetFinite$ or $\Distributions $ also the converse
  inequality holds, and thus we obtain the result.  We proceed
  separately for the two functors.

  \medskip
  
\noindent\emph{Finite powerset.}
  Let us denote $d = \sup_i d_i$. We have to show that 
  \begin{center}
    $\LiftedMetric{\PowersetFinite}{d} \leq 
    \sup_i d_i^{\PowersetFinite}$
  \end{center}
  Let $X_1, X_2 \in \PowersetFinite(X)$ be finite subsets of
  $X$. Since $X_1$ and $X_2$ are finite and $d = \sup d_i$, for any
  $\epsilon > 0$ we can find an $i \in \N$ such that for any
  $x_1 \in X_1$, $x_2 \in X_2$ and all $j \geq i$
  \begin{align*}
    d(x_1, x_2) - d_j(x_1, x_2) \leq \epsilon\,.
  \end{align*}
According to the definition of the Wasserstein lifting $\PowersetFinite$ we get for all $j \geq i$:
\begin{align*}
	\Wasserstein{\PowersetFinite}{d}(X_1,X_2) &= \inf \set{\max_{(x_1,x_2) \in W} d(x_1,x_2) \mid W \in \Couplings{\PowersetFinite}(X_1, X_2)}\\
&\leq \inf \set{\max_{(x_1,x_2) \in W} (d_j(x_1,x_2) + \epsilon) \mid W \in \Couplings{\PowersetFinite}(X_1, X_2))}\\
     &= \inf \set{\max_{(x_1,x_2) \in W} d_j(x_1,x_2) \mid W \in \Couplings{\PowersetFinite}(X_1, X_2)} + \epsilon= \Wasserstein{\PowersetFinite}{d_j}(X_1,X_2) + \epsilon
\end{align*}
  
\noindent Therefore, $\LiftedMetric{\PowersetFinite}{d}(X_1,X_2) \leq \sup_i d_i^{\PowersetFinite}(X_1,X_2) + \epsilon$. Given that $\epsilon$ can be arbitrarily small, we deduce that indeed $\LiftedMetric{\PowersetFinite}{d}(X_1,X_2) \leq \sup_i d_i^{\PowersetFinite}(X_1,X_2)$, as desired.

  \medskip

  \noindent
  \emph{Finitely supported distributions.}
  Let us denote $d = \sup_i d_i$. We have to show that 
  \begin{center}
    $d^{\Distributions } \leq 
    \sup_i d_i^{\Distributions }$
  \end{center}
  Let $P_1, P_2 \in \Distributions X$ and let $X_1$, $X_2$ be the corresponding finite supports of $P_1$ and $P_2$, namely $X_i = \{ x \in X \mid P_i(X) > 0 \}$. As before, since $X_1$
  and $X_2$ are finite and $d = \sup d_i$, for any $\epsilon > 0$ we
  can find an $i \in \N$ such that for any $x_1 \in X_1$, $x_2
  \in X_2$ and $j \geq i$
  \begin{align*}
    d(x_1, x_2) - d_j(x_1, x_2) \leq \epsilon\,.
  \end{align*}
  Using the definition of the Wasserstein lifting of $\Distributions $,
  we get for all $j \geq i$:
\begin{align*}
	\Wasserstein{\Distributions}{d}(X_1,X_2) = \inf \set{\sum_{x_1,x_2 \in X} d(x_1,x_2) \cdot P(x_1, x_2) \mid P \in \Couplings{\Distributions }(P_1, P_2)}
\end{align*}
  since for any $(x_1, x_2) \in X \times X$, whenever $(x_1, x_2) \not\in X_1\times X_2$ necessarily $P(x_1,x_2)=0$
\begin{align*}
	&= \inf \set{\sum_{(x_1,x_2) \in X_1 \times X_2} d(x_1,x_2) \cdot P(x_1, x_2) \mid P \in \Couplings{\Distributions }(P_1, P_2)}\\
	&\leq \inf \set{\sum_{(x_1,x_2) \in X_1 \times X_2} (d_j(x_1,x_2) + \epsilon) \cdot P(x_1, x_2) \mid P \in \Couplings{\Distributions }(P_1, P_2)}\\
	&= \inf \set{\sum_{(x_1,x_2) \in X_1 \times X_2} d_j(x_1,x_2) \cdot P(x_1, x_2)  + \epsilon \cdot \sum_{x_1,x_2 \in X} P(x_1,x_2) \mid P \in \Couplings{\Distributions }(P_1, P_2)}\\
	&= \inf \set{\sum_{(x_1,x_2) \in X_1 \times X_2} d_j(x_1,x_2) \cdot P(x_1, x_2)  + \epsilon \mid P \in \Couplings{\Distributions }(P_1, P_2)}\\
	&= \inf \set{\sum_{(x_1,x_2) \in X_1 \times X_2} d_j(x_1,x_2) \cdot P(x_1, x_2) \mid P \in \Couplings{\Distributions }(P_1, P_2)}  + \epsilon\\
	&= \inf \set{\sum_{x_1,x_2 \in X} d_j(x_1,x_2) \cdot P(x_1, x_2) \mid P \in \Couplings{\Distributions }(P_1, P_2)}  + \epsilon
	= \Wasserstein{\Distributions}{d_j}(X_1,X_2) + \epsilon
\end{align*}
  
\noindent Therefore, $d^{\Distributions }(X_1,X_2) \leq \sup_i d_i^{\Distributions }(X_1,X_2) + \epsilon$. Given that $\epsilon$ can be arbitrarily small, we deduce that indeed $d^{\Distributions }(X_1,X_2) \leq \sup_i d_i^{\Distributions }(X_1,X_2)$, as desired.
\end{proof}

\restate{thm:comp-dist}
\begin{proof}
The chain $e_i$ of metrics is the same as the one constructed in the proof of \theoremref{thm:final-coalgebra}. In this proof we have shown that all $f\colon (X,e_i)\nonexpansiveTo (\Omega,d_i)$ are nonexpansive. Here we show that they are also isometries. For $i=0$ this is true: for $x,y\in X$ $d_0(f(x),f(y)) = 0 = e_0(x,y)$. Now assume that $f\colon (X,e_i)\nonexpansiveTo (\Omega,d_i)$ is an isometry, which implies ($\LiftedFunctor{F}$ preserves isometries) that $\LiftedFunctor{F}f\colon (FX,\LiftedMetric{F}{e_i})\nonexpansiveTo (F\Omega,\LiftedMetric{F}{d_i})$ is an isometry. Hence for $x,y\in X$ we have
\begin{align*}
	d_{i+1}(f(x),f(y)) & = \LiftedMetric{F}{d_i}(\kappa(f(x)),\kappa(f(y))) = \LiftedMetric{F}{d_i}(Ff(\alpha(x)),Ff(\alpha(y))) \\
		& = \LiftedMetric{F}{e_i}(\alpha(x),\alpha(y)) = e_{i+1}(x,y) 
\end{align*}
For a limit ordinal $j$ we have 
\begin{equation*}
	d_j(f(x),f(y)) = \sup_{i<j} d_i(f(x),f(y)) = \sup_{i<j} e_i(x,y) = e_j(x,y)
\end{equation*}
We know that $d_\theta$ is a fixed-point, i.e. we have $d_\theta = d_\theta^F\circ (\kappa\times \kappa)$. Then $e_\theta$ must also be a fixed-point ($e_\theta = e_\theta^F\circ (\alpha\times \alpha)$), since:
\begin{align*}
	e_\theta(x,y) & = d_\theta(f(x),f(y)) = d_\theta^F(\kappa(f(x)),\kappa(f(y))) = d_\theta^F(Ff(\alpha(x)),Ff(\alpha(y))) \\
		& = e_\theta^F(\alpha(x),\alpha(y))
\end{align*}
using again the fact that $Ff$ is an isometry. Hence $\zeta \le \theta$, i.e., the chain $e_i$ might converge earlier and $d_\kappa(x,y) = d_\theta(f(x),f(y)) = e_\theta(x,y) = e_\zeta(x,y)$.
\end{proof}

\restate{thm:d-omega-is-metric}
\begin{proof}
We construct a series of metrics $e_i\colon W_i \times W_i \to \reals$ for $i\in\mathsf{Ord}$, as follows: $e_0 = 0\colon \one \times \one \to \reals$ is the (unique!) zero metric on $\one$, $e_{i+1} := \LiftedMetric{F}{e_i}$ and if $j$ is a limit ordinal we define $e_j := \sup_{i<j} e_i\circ (p_{j,i}\times p_{j,i}) \text{ }$.
Since the functor preserves metrics $e_{i+1}$ is a metric if $e_i$ is. Given a limit ordinal $j$ we can easily check that $e_j$ is a pseudometric provided that all the $e_i$ with $i < j$ are pseudometrics. To see that $e_j$ is also a metric when all $e_i$ with $i<j$ are metrics we proceed as follows: Suppose $e_j(x,y) = 0$ for some $x,y \in W_j$, then we know that for all $i$ we must have $p_{j,i}(x) = p_{j,i}(y)$ because the $e_i$ are metrics. Since the cone $(W_j \stackrel{p_{j,i}}{\to} W_i)_{i < j}$ is by definition a limit in $\Set$ we can now conclude that $x=y$. This is due to the universal property of the limit: If $x \not =y$ then for the cone $(\set{x,y} \stackrel{f_i}{\to} W_i)_{i<j}$ with $f_i(x) = p_{i,j}(x) = f_i(y)$ there would have to be a unique function $u\colon \set{x,y} \to W_j$ satisfying $p_{i,j}\circ u = f_i$. However, for example $u, u'\colon \set{x,y} \to W_j$ where $u(x) = u(y) = x$ and $u'(x) = u'(y) = y$ are distinct functions satisfying this commutativity which is a contradiction to the uniqueness.

Using the metrics $e_i$ we now consider the connection morphisms $p^i := p_{\lambda,i}\colon \Omega\to W_i$ and proceed by showing that each $p^i$ is an isometry $(\Omega,d_i) \nonexpansiveTo (W_i,e_i)$. By definition this holds for  $d_0$ and $e_0$ (both are constantly zero). For $i+1$ we recall that $\kappa = p_{\lambda+1, \lambda}^{-1}$. Hence we have by properties of the connection morphisms $Fp^i = Fp_{\lambda,i} = p_{\lambda+1,i+1} = p_{\lambda,i+1} \circ p_{\lambda+1,\lambda} = p^{i+1} \circ \kappa^{-1}$ and thus $Fp^i \circ \kappa = p^{i+1}$. Since by the induction hypothesis, $p^i\colon (\Omega,d_i)\nonexpansiveTo (W_i,e_i)$ is an isometry, the fact that isometries are preserved by $\LiftedFunctor{F}$ implies that $\LiftedFunctor{F}p^i\colon (F\Omega,\LiftedMetric{F}{d_i})\nonexpansiveTo (W_{i+1},e_{i+1})$ is an isometry. Furthermore $\kappa\colon (\Omega,d_{i+1})\nonexpansiveTo (F\Omega,\LiftedMetric{F}{d_i})$ is an isometry. Hence also their composition $p^{i+1} = Fp^i \circ \kappa$ is an isometry. For a limit ordinal $j$ we calculate for $x,y \in \Omega$
\begin{align*}
	e_j(p^j(x),p^j(y)) & = \sup_{i<j} e_i((p_{j,i} \circ p^j)(x),(p_{j,i} \circ p^j)(y)) = \sup_{i<j} e_i(p^i(x),p^i(y)) \\
    & = \sup_{i<j} d_i(x,y) = d_j(x,y)
\end{align*}
and thus also $p^j\colon (\Omega,d_j)\to (W_j,e_j)$ is an isometry,.

We assume that $\lambda\ge \theta$, otherwise set $\lambda = \theta$ (if the final chain converges in $\theta$ steps it also converges for all larger ordinals), thus $d_\lambda = d_\theta$. Now let $x,y\in\Omega$ with $d_\theta(x,y) = d_\lambda(x,y) = 0$. This implies $e_i(p^i(x),p^i(y)) = 0$ for all ordinals $i\le\lambda$. Since all $e_i$ are metrics, we infer that $p^i(x) = p^i(y)$ for all ordinals $i$. With the same reasoning as above (where we proved that $e_j$ is a metric for limit ordinals $j$) this implies that $x$ and $y$ are equal. 
\end{proof}

%%% Local Variables: 
%%% mode: latex
%%% TeX-master: "coalgebra-metrics-lipics"
%%% End: 